"Normalized insured losses caused by windstorms in Quebec and Ontario, Canada, in the period 2010–2020"

Mohammad Hadavi*[1], Lutong Sun[1], Djordje Romanic[1]

[1] Department of Atmospheric and Oceanic Sciences, Faculty of Science, McGill University

* Corresponding author.

Department of Atmospheric and Oceanic Sciences, Faculty of Science, McGill University, Burnside Hall, Office 835, 805 Sherbrooke Street West, Montreal, Quebec H3A 0B9, Canada.

E-mail: mohammad.hadavi@mail.mcgill.ca.     Phone: +1 (514) 443 7747

E-mail address of authors:

mohammad.hadavi@mail.mcgill.ca

lutong.sun@mail.mcgill.ca

djordje.romanic@mcgill.ca

Abstract

Severe windstorms pose threats to people, human-made structures, and the environment. An investigation of insured losses caused by windstorms is a multipurpose study that serves to advance the sustainability and resilience of modern communities. The present study proposes a systematic analysis of insured losses imposed by different types of windstorms in two Canadian provinces, Ontario (ON) and Quebec (QC), during the period 2010–2020. Actual wind damage data from the Canadian insurance market were considered in this study. Our calculations show that ON and QC received half of all wind catastrophes across Canada, and nearly three-quarters of these catastrophes were wind-related ones. The total windstorm loss of over CA$4.5 billion was not evenly distributed between QC and ON, but rather had a QC:ON ratio of 1:3.3. We attributed this discrepancy in the inflicted damage between two provinces to the predominantly eastward and northeastward trend of storm trajectories and the higher density of wealth and population in ON. Convective storms were the most devastating wind type comprising nearly 60% and 70% of the total number of events and associated damage, respectively, in the two provinces. Finally, tornadoes had the highest average loss per event.

Graphical abstract

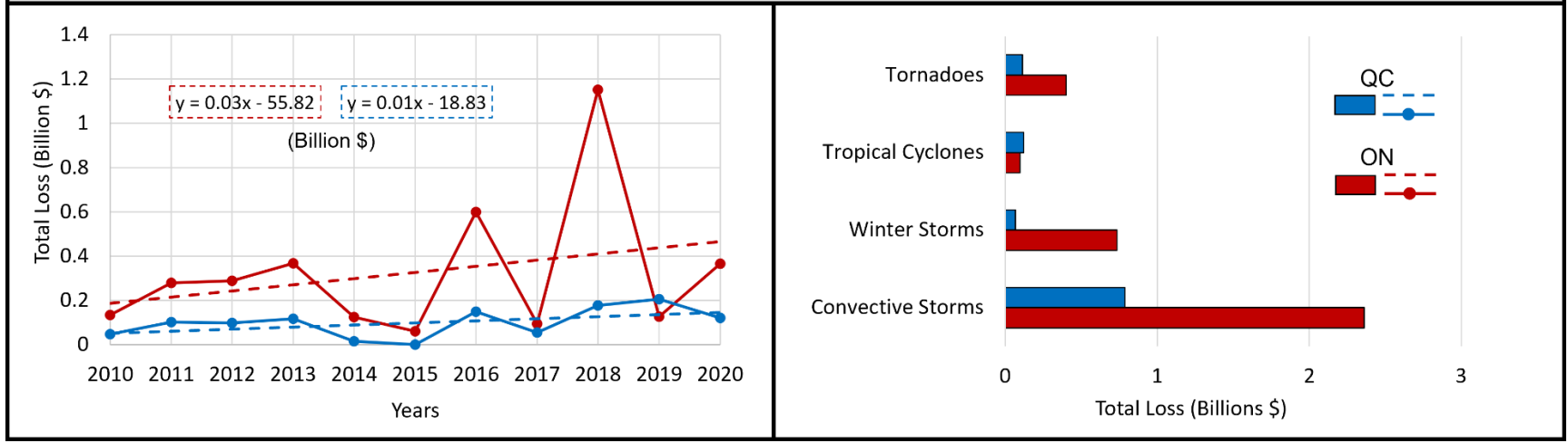

Highlights

- Insurance data were used to study windstorm damage in Ontario (ON) and Quebec (QC)
- ON and QC account for half of all wind catastrophes across Canada in 2010–2020
- 73.6% of total catastrophes in ON and QC were wind-related events
- Windstorms imposed more than CA$4.5 billion loss with the ratio of 1:3.3 in QC:ON
- Convective storms were the most disastrous wind type causing ~70% of the total loss

# 1  Introduction

Globally, windstorms have been the deadliest type of weather-related disaster by causing more than 242,000 deaths, and they are also the second most frequent natural hazard by accounting for 28% of total events in 1995–2015 (United Nations, 2015). Violent windstorms pose a serious threat to people and man-made structures, such as power line systems (Dai et al., 2017), buildings (Simmons et al., 2015), bridges (Burlando et al., 2020), and ports (Solari et al., 2012), to name a few. Moreover, literature suggests that strong wind episodes are expected to become more frequent and intense because of climate change (Diffenbaugh et al., 2013; McBean, 2004; Seeley & Romps, 2015). With these predictions in mind, it is logical to expect that wind disaster assessment studies will play one of the central roles of improving the sustainability and resilience of modern communities to natural disasters. The issue of extreme wind speeds is additionally highlighted in the Canada's Changing Climate Report (CCCR, 2019) by noticing that although extreme winds are relevant for sustainable wind energy production and resilient buildings, there are limited analysis on the damage and mechanisms of extreme winds in Canada.

The financial burden to individuals and communities caused by different types of severe winds has already received moderate attention in the scientific community. In terms of tornado damage, Romanic et al. (2016) developed a Monte Carlo model for loss assessment and applied it for the state of Oklahoma, United States (US). Subsequently, Refan et al. (2020) introduced a revised version of this model by improving fragility modelling, tornado flow field dynamics, and also extended the exposure map to Kansas, US. The introduction of fragility curves instead of vulnerability curves and more realistic tornado flow field resulted in a reduction of economic losses compared to Romanic et al. (2016). Huang et al. (2021) developed a preferred prediction model for the tornado occurrence rate for Canada. By incorporating annual average cloud-to-ground lightning flash density and annual average thunderstorm days as covariates in the prediction model, they showed that tornado occurrence in Canada is associated with large overdispersion. Gumaro et al. (2022) conducted vulnerability assessments to identify strategies for improving the resilience of buildings against the severe tropical cyclones which have caused significant damages in the Philippines. Proposing a simulation framework, Gu et al. (2021) analyzed wind damage to urban trees which may result in direct and indirect economic losses. Pant and Cha (2019a, 2019b) focused on hurricane risk assessment and hurricane loss, respectively, across the coastal regions of the US and they found that hurricane risk and associated losses are expected to increase in future climate compared to the present conditions for all studied locations and climate change scenarios. Though the intensity of risk escalation showed pronounced spatial variability, the highest risk was found towards the northeast side of the US.

Evaluating potential trends in natural catastrophe losses requires compensating for changes in asset values and exposures over time (Miller et al., 2008). The population and the properties at risk both accumulate as time goes on, so a normalization method should be employed to take into account the impacts of inflation, enhancing population, and increasing density of wealth on the total amount of loss (Pielke, 2021). Together with proposing a normalization method to compare damage from different historical periods, Pielke and Landsea (1998) studied the US hurricanes in the period of 1925–1995 and when the annual average normalized damage was about US$5.8 billion. They showed that the trend of increasing damage identified by most concurrent studies at that time disappears as a result of using the newly (at that time) proposed normalization method that considered the factors such as inflation, population trends, and wealth differences between different time periods. After that, a considerable number of normalization studies have emerged

seeking to remove the influences of societal and economic changes from the loss time series of different disasters. A majority of these studies focused on floods and storms (Pielke, 2021).

Zhang et al. (2009) examined direct economic losses and casualties associated with tropical cyclones over China in 1983–2006. The authors found that the direct economic losses had a significant upward trend in this period, while no trend was detected for the losses scaled with the total annual Gross Domestic Product (GDP) of China and the annual GDP per capita. This finding implies that the increasing trend in direct economic losses was mainly due to Chinese economic development. Assessing the economic losses of four selected ports in China, Zhang et al. (2020) evaluated the economic impact of typhoon-induced wind disasters on port operations. Their result showed that the loss to the shippers has been the major component of overall loss. Nordhaus (2010) examined the economic impacts of the US hurricanes by using a regression-based approach and estimated that the normalized damage would be sharply enhanced for US$10 billion at 2005 incomes (0.08 percent of GDP) because of stronger hurricanes caused by global warming. However, performing a similar statistical analysis, Bouwer and Wouter Botzen (2011) quantitatively re-evaluated the results of Nordhaus (2010) and found that the upward trend disappears by using more accurate data series adjusted for local changes in exposure of assets over time. These contrasting findings can also be seen in more recent studies; for instance, Grinsted et al. (2019) addressed climatic trends of the US hurricane damage in 1900–2018 using the area of total destruction and found that hurricanes are becoming more damaging. On the other hand, Martinez (2020) did not find such a trend in the US normalized hurricane losses that account for changes in building costs. These contradictory trends are not surprising, as they can be attributed to different approaches to normalization (Pielke, 2021). In general, most weather- and climate-related normalization studies in 1998–2020 ($\simeq 73\%$) found neither upward nor downward trends

for normalized losses in the most prolonged period examined by each study, while they might show trends for sub-periods of their time series (Pielke, 2021).

A number of risk assessment studies have looked at disaster losses on a global scale (Cardona et al., 2014; Pielke, 2019; Visser et al., 2014; Yamin et al., 2014). However, because the global scale losses seem to be dominated by the losses induced in wealthy regions (Mohleji & Pielke Jr, 2014; Watts et al., 2019), more studies have concentrated on the specific regions of the world. Considering wind-related disasters only, extratropical storms were investigated in Switzerland (Stucki et al., 2014) and Europe (Barredo, 2010), while tropical cyclones were studied over China (Chen et al., 2018; Fischer et al., 2015; Ye & Fang, 2018), India (Raghavan & Rajesh, 2003), and Latin America and the Caribbean (Pielke Jr et al., 2003). The continental US has been thoroughly investigated by a majority of normalization studies regarding different windstorms, such as tropical cyclones (Estrada et al., 2015; Klotzbach et al., 2018; Pielke Jr et al., 2008; Schmidt et al., 2009; Weinkle et al., 2018) and tornadoes (Boruff et al., 2003; Brooks & Doswell III, 2001; Simmons et al., 2013). In regard to damaging thunderstorm activities in the US, Changnon (2001) reported that 892 catastrophic events resulted in US$87 billion loss during 1949–1998. Sander et al. (2013) exhibited that the rising trend of thunderstorm-related losses over time is consistent with the expected impacts of anthropogenic climate change on the forcing of convective storms.

Based on the detailed literature review conducted by Pielke (2021), none of the previous studies considered Canada in terms of normalized damage losses caused by windstorms. While severe wind hazards are most frequent in the US (Mohleji & Pielke Jr, 2014), they have also been causing significant damage in highly populated areas in Canada (Harrison et al., 2015). Hence, by addressing this research gap, the present study aims at a systematic analysis of insured losses imposed by different types of windstorms in two provinces of Canada—Ontario (ON) and Quebec (QC). The target period is from 2010 through 2020, and the source data are the actual insurance losses caused by various windstorms. These two specific provinces were selected because our analysis has shown that almost 50% of all catastrophic wind events across Canada in 2010–2020 occurred in ON and QC.

The present study investigates damage induced by synoptic and non-synoptic wind systems. We also provide a comparative analysis of damage inflicted by different types of windstorms and discuss them in terms of geographical and socio-economic differences between ON and QC. In terms of wind engineering practice, the current building codes and recommendations in Canada are predominantly tailored towards the severe but large-scale atmospheric boundary layer winds (i.e., commonly known as synoptic winds in the wind engineering terms) associated with hurricanes and extratropical cyclones (Davenport, 1961). Nevertheless, our research demonstrates that the majority of normalized wind losses in ON and QC are caused by convective storms (i.e., non-synoptic winds) rather than synoptic winds. Finally, it should be remarked that the high density of population and wealth in urban areas makes these regions the most vulnerable to extreme weather (Henstra, 2012; Hunt, 2004). Therefore, the results of this research serve towards improving the sustainability and resilience of Canadian cities and communities against one of the deadliest types of disasters, named windstorms.

## 2 Data and methodology

### 2.1 CatIQ database

The present research employs the dataset hosted and maintained by a Toronto-based insurance firm Catastrophe Indices and Quantification Inc. (CatIQ, 2021). This database contains detailed analytical and meteorological information on Canadian natural and human-made catastrophes, and it is widely recognized as the most reliable source of catastrophe loss information in Canada. CatIQ is a subsidiary of Zurich-based PERILS AG and was established in 2014 with the support of most of the Canadian insurance and reinsurance industry. This database contains loss information collected directly from insurers, along with meteorological and damage details for all catastrophes in Canada since 2008. CatIQ's data are directly collected from insurance companies writing business in Canada. CatIQ aggregates the submitted data by different insurance companies and then extrapolates all collected data to the industry level using market share information. The calculated industry event loss is made available through the online CatIQ Platform at https://public.catiq.com/.

Most previous studies (Brooks & Doswell III, 2001; Nordhaus, 2010; Pielke Jr et al., 2008; Schmidt et al., 2009) considered the total economic losses, i.e., the sum of insured and uninsured losses, that is typically estimated as multiples of insured losses. This approach tends to reduce the reliability of the estimated damage because the multiplier factor is often difficult to assess accurately (Barthel & Neumayer, 2012). This issue was further discussed by Romanic et al. (2016). On the other hand, while the reported insured losses have higher fidelity, the numbers are not providing the total loss caused by a catastrophe. While acknowledging the positives and shortcomings of both methods, our study is based on the insured losses. With that in mind, the "insured" term is mostly omitted hereafter because all reported and discussed data are insured losses.

In this paper, different wind-related catastrophes have been classified into the following four categories: (1) Convective storms: including deep convection, squall lines, downbursts, and other thunderstorm types that are not observed in winter and snowy conditions; (2) Winter storms: including ice storms and frozen precipitation also characterized by strong winds; (3) Tropical cyclones: including hurricanes and tropical storms; and (4) Tornadoes. Fig. 1 depicts a few examples of wind-related catastrophes in ON and QC caused by these four wind categories.

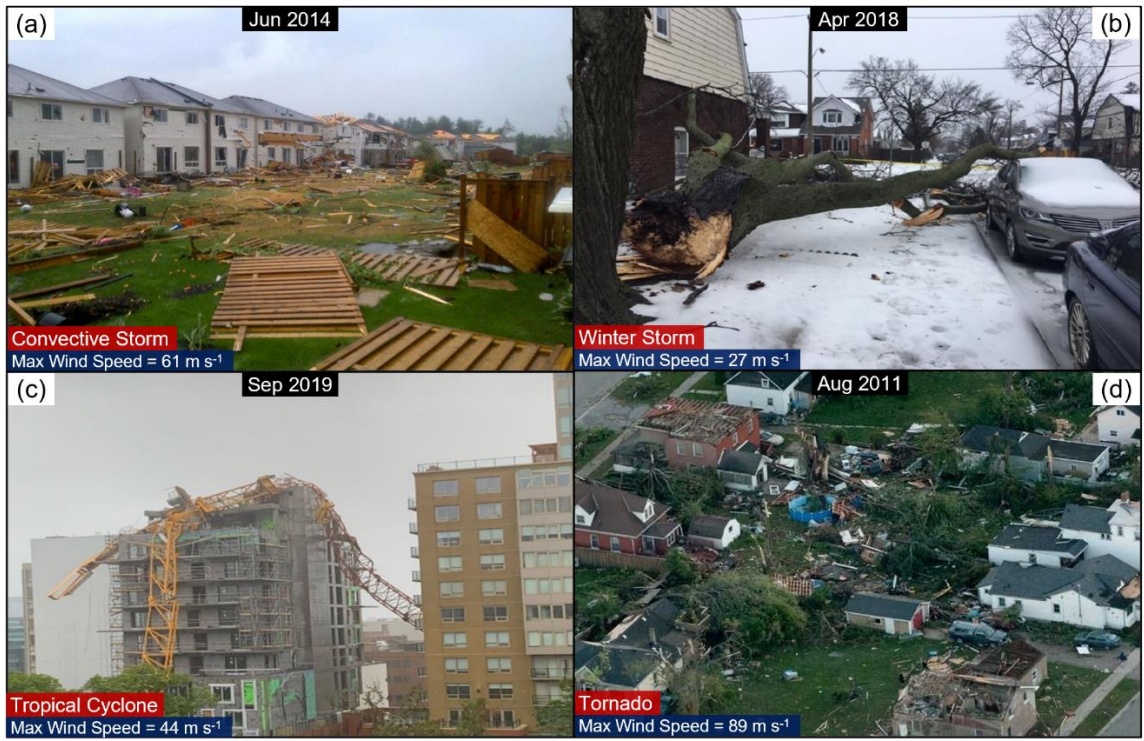

Fig. 1. Photographs of wind-related catastrophes. (a) Southern Ontario Severe Storms (Global News, 2014), (b) Southern Ontario and Quebec Ice Storm (Global News, 2018), (c) Post-Tropical Cyclone Dorian (Benjamin, 2019), and (d) Goderich Tornado (Rawlings, 2011). An estimate of the maximum wind gust is also included in the figure. The maximum wind gusts come from the Environment & Climate Change Canada (ECCC) storm summary reports.

In the CatIQ database as well as the present study, a catastrophe is defined as an event that caused more than CA$25 million in losses for the industry, affected multiple policyholders and insurers. Hereafter, all insured losses in this paper are expressed in Canadian dollars; therefore, the abbreviation CA$ is replaced with $. Based on this definition, there were 47 wind-related catastrophes in the period 2010–2020 in ON and QC. Catastrophes in the database were reported based on a combination of different perils, not a single one, as the inflicted damage often cannot be uniquely associated to a single peril. For example, in a case of a severe thunderstorm, damage can be caused by one, a few, or all of the following high-impact weather perils: severe straight-line winds, tornadoes, hail, intense precipitation associated with flooding, and lightning. During a post-disaster damage survey, it is often difficult to precisely separate the amount of damage caused by straight-line winds (e.g., downbursts) versus tornado(s) versus flooding that can be triggered by a combination of intense precipitation and previously inflicted damage by severe winds. Moreover, Kelly et al. (1985) discussed that people often tend to report only the most spectacular events associated with thunderstorms and omit other phenomena. For instance, if a tornado outbreak occurred over a region, then the majority of damage is often assigned to tornadoes, while in reality some of the damage might have been caused by other types of thunderstorm winds. To account for that challenge, each catastrophic loss event in the CatIQ database is tagged with up to four perils associated with the observed damage. The perils are listed in descending order starting from the one that likely caused the most significant loss. To study the wind-related damage specifically, all records that contained a windstorm tag were examined one by one, resulting in a total of 47 catastrophic events. Subsequently, 8 of these events were excluded from further analysis because the wind had a low impact and was ranked as the last affecting peril for that event. Therefore, a total of 39 events entered the final analysis.

## 2.2 Normalization process

To take into account the wealth accumulation over time, obtained data have been adjusted to scale up the loss from past disasters. The conventional approach for normalizing natural disasters (Pielke et al., 1999; Pielke & Landsea, 1998) assumes three correction factors: inflation (i.e., the change in products prices), population, and wealth. Inflation is accounted for by using the Gross Domestic Product (GDP) deflator, viz.:

$$\text{Normalized Loss}_t^s = \text{Loss}_t \times \left(\frac{\text{GDP deflator}_s}{\text{GDP deflator}_t}\right) \times \left(\frac{\text{Population}_s}{\text{Population}_t}\right) \times \left(\frac{\text{Wealth per capita}_s}{\text{Wealth per capita}_t}\right), \quad (1)$$

where $s$ is the reference year (2019) for normalization, and $t$ is the catastrophic event's year.

Later studies were dedicated to finding the more appropriate proxies that can better describe each factor under different circumstances (Barthel & Neumayer, 2012; Neumayer & Barthel, 2011). For global analysis, GDP per capita has been used as a proxy to represent wealth at larger scales. Moreover, Barthel and Neumayer (2012) adjusted the conventional normalization methodology represented in Eq. (1) by adding an additional factor to control for changes in the insurance penetration (also known as "insurance take up rates" or "insurance gaps"), whose level provides an indicator of the relative size and importance of insurance in the domestic economy (OECD, 2016). Finally, considering that the product of population and GDP per capita equals total GDP, Barthel and Neumayer (2012) presented an improved normalization method that has been used in the present study to normalize wind loss data:

$$\text{Normalized Loss}_t^s = \text{Loss}_t \times \left(\frac{\text{GDP deflator}_s}{\text{GDP deflator}_t}\right) \times \left(\frac{\text{GDP}_s}{\text{GDP}_t}\right) \times \left(\frac{\text{Ins. Penetration}_s}{\text{Ins. Penetration}_t}\right). \quad (2)$$

This study employs The World Bank (2020), Statistics Canada (2020), and OECD (2020) to gather required data for GDP deflator, total GDP, and insurance penetration for Canada, respectively.

# 3 Results and discussion

## 3.1 Wind catastrophes in Ontario and Quebec

In terms of overall significance and one of the highlights of this research, we show that 64.2% of total catastrophic losses in ON and QC between 2010 and 2020 were caused by windstorms (Fig. 2). Provincially, wind-related catastrophes were responsible for 65.7% and 59.6% of total losses in ON and QC, respectively. Moreover, there were only 14 catastrophes in which the main peril was not wind but flooding, thus implying that 73.6% of all catastrophes in ON and QC between 2010 and 2020 were wind-related events. Furthermore, the analysis of loss data across all of Canada in the period 2010–2020 (not shown) demonstrated that almost half of all wind-related catastrophes occurred in ON and QC. Hence, the present study focuses on ON and QC.

Fig. 2 also depicts the total number of wind-related catastrophes and the overall insured loss in the period 2010–2020 for ON and QC, separately. The two studied provinces show meaningful differences in the loss results. Whereas the number of occurred catastrophes are not significantly different—3 and 2.18 wind events per year in ON and QC, respectively—the total imposed loss in ON is more than three times that of QC. The yearly analysis in Fig. 3 shows that 2019 was the only year in which QC experienced a slightly higher insured loss than ON due to a strong convective storm event that occurred in October 2019 and significantly affected the area around Montreal. Excluding this year, however, insured losses were higher in ON than QC in all years, particularly in 2016 and 2018. This finding will be discussed in more detail later in this study.

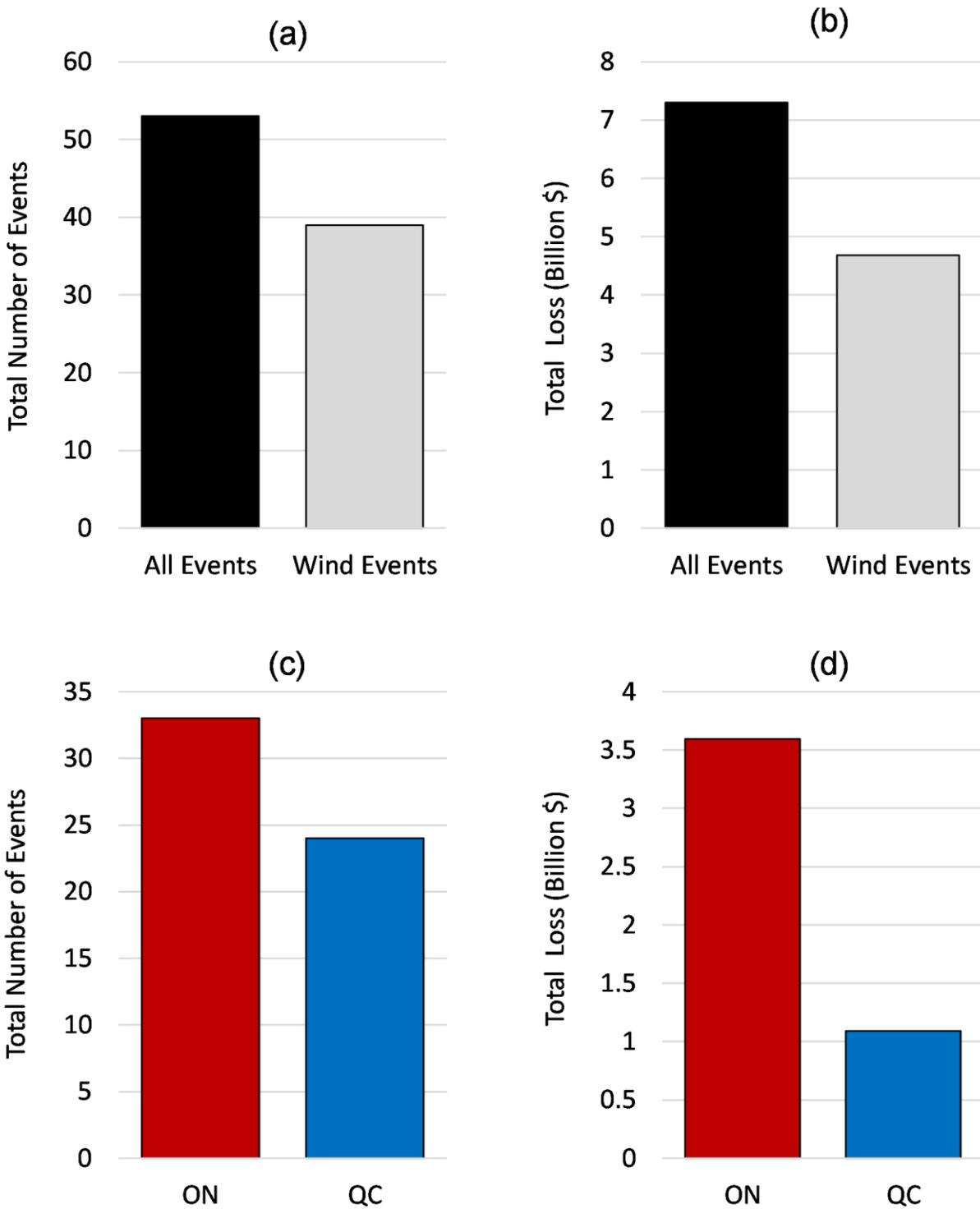

Fig. 2. (a) total number of all catastrophes and wind-related catastrophes and (b) their total imposed loss in both provinces combined in the period 2010–2020. (c) total number of catastrophic wind events and (d) their total wind damage in ON and QC in the investigated period.

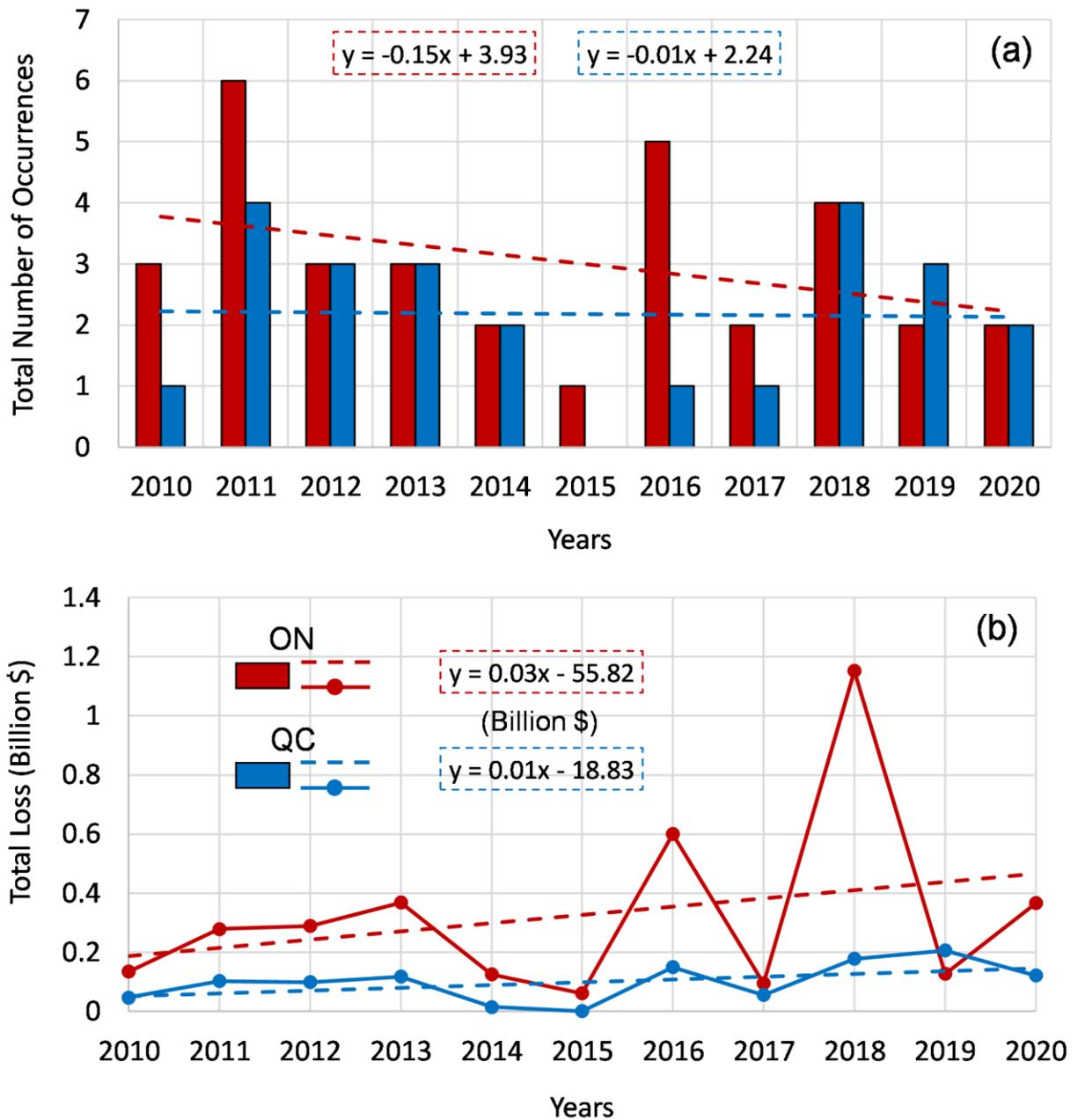

Fig. 3. (a) Total number of catastrophic wind events per year and (b) their annual loss in ON and QC in the period 2010–2020. The straight lines are linear trends with the trend equations provided in the plot.

The observed discrepancy between the total insured losses in two provinces—1:3.3 in terms of QC:ON—is attributed to the following two factors. First, many of the weather systems that resulted in wind-related catastrophes in QC approached the province from southern ON, particularly in the case of convective storms. Fig. 4 presents trajectories of weather systems causing wind catastrophes in ON and QC. The trajectories were obtained by individually analyzing synoptic charts at the times of identified wind events. The eastward and northeastward trends of storm trajectories resulted in more wind damage in ON because the intensity of most storm systems decreased by the time they reached QC. Fig. 4 also depicts that the major damage was situated in southern parts of ON and QC.

Second, exposure to natural disasters, including wind catastrophes, is higher in ON due to higher population density and more wealthy regions (i.e., very high exposure) in the Greater Toronto Area (GTA). As discussed in Section 2, while loss normalization adjusts for the increasing disaster losses in a time series, it does not take into account (and it should not) the differences in the wealth level across the provinces. Namely, a given catastrophic wind event will cause a substantially different level of damage depending on the population density and the value of assets at risk of the exposed region. Therefore, exploring the average loss per event in ON and QC can better clarify the role of demographic data in wind damage analysis. As demonstrated in Fig. 5, the average loss per event in ON is higher than QC in most years in the studied period. The year 2016 was an exception to this pattern as a result of the pronounced difference in the number of wind catastrophes that occurred in ON (=5) and QC (=1). The average loss per event in the period 2010–2020 is almost $109 and $45 million in ON and QC, respectively, owing to the above-mentioned greater density of wealth and population in ON, primarily the GTA. Our data show that the ratio QC:ON in terms of an average wind damage per event is 1:2.4.

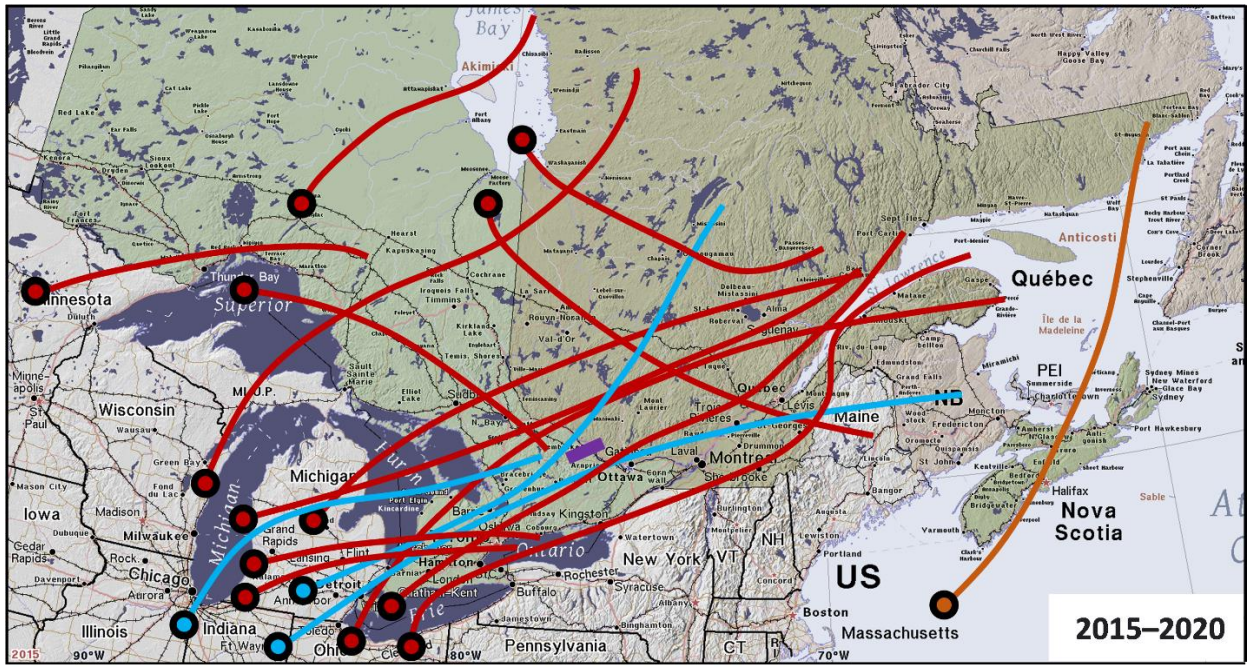

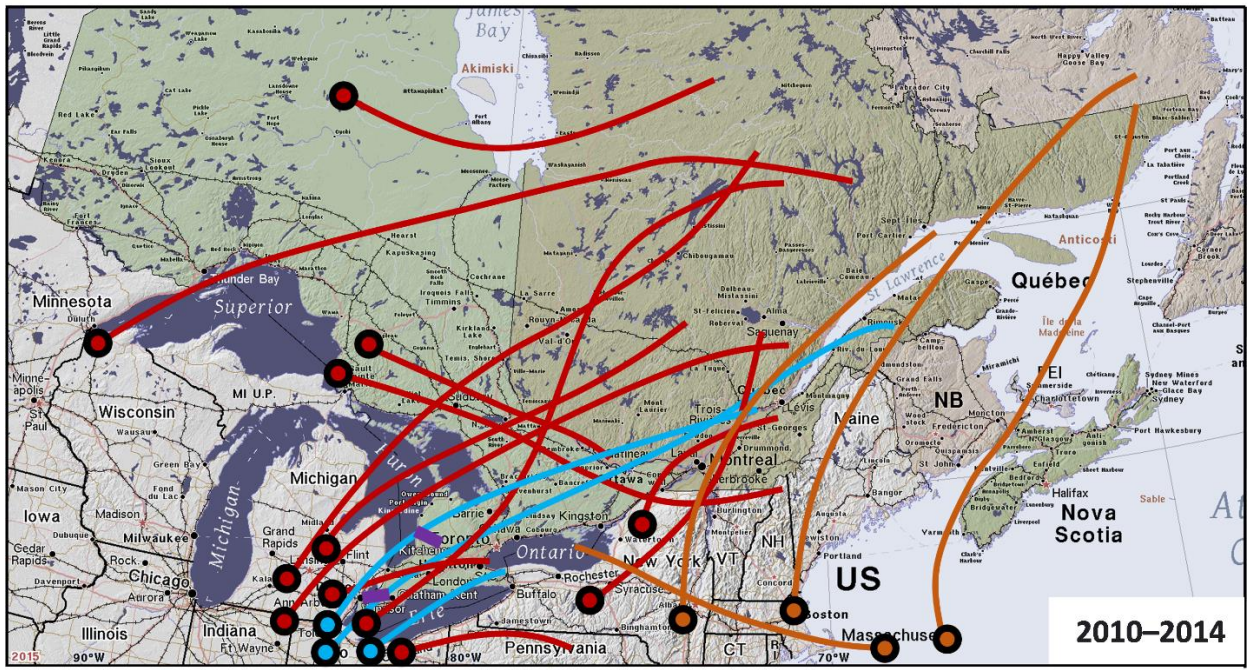

Fig. 4. Trajectories of weather systems causing wind-related catastrophes in ON and QC in the period 2010–2020. For convenience, the figure is split into the periods 2010–2014 and 2015–2020 in order to make the trajectories more visible. The circles indicate the up-trajectory positions, i.e., the weather systems moved from the circles towards the other end of the line.

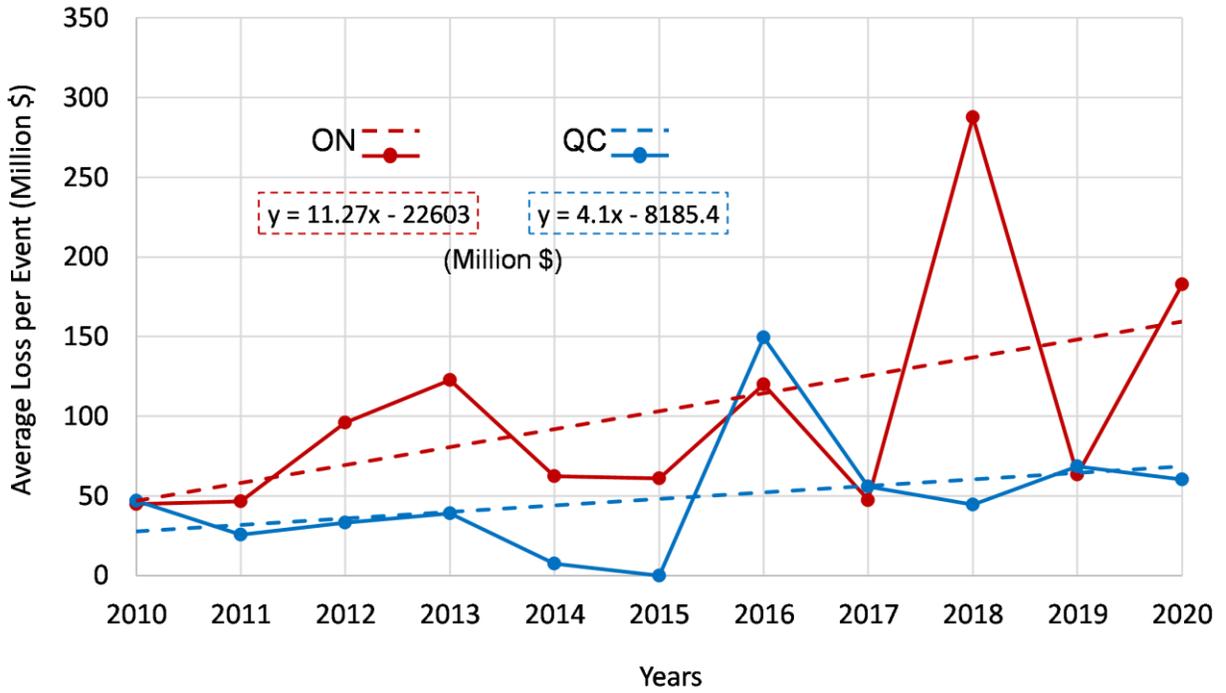

Fig. 5. The yearly trend of average loss per event for ON and QC in the period 2010–2020. The straight lines are linear trends with the trend equations provided in the plot.

## 3.2 Annual analysis of wind catastrophes

The normalized data depicted in Fig. 3 indicate that insured loss peaked in 2018. While the province of ON only experienced four catastrophic wind events in that year, it financially sustained a heavy loss of about $1.15 billion. Table 1 elucidates this observation by sorting the top five most destructive catastrophes across QC and ON. Notice that three of them occurred in 2018. The highest-ranked wind catastrophe was a devastating convective storm causing almost half of the above-mentioned wind damage by affecting the GTA and Hamilton. The wind gusts in this event exceeded 35 m s$^{-1}$ and the severe winds that sustained for several hours caused widespread damage. However, this event was very peculiar because the strongest wind gusts were observed after the passage of deep thunderstorms. Further meteorological analysis of this event (Weatherlogics, 2018) suggests that a cold front associated with convective storms as well as a trough of low pressure that passed the GTA and Hamilton region after the convective storms forced extremely strong winds from aloft (near 850 hPa) to the surface. On the other hand, only some parts of southern QC were affected by this windstorm, hence causing smaller wind damage in QC than ON (the damage ratio of almost 1:14). Overall, 2018 is a clear example of the situation in which the same number of catastrophes caused significantly different insured losses between the two provinces (the ratio of 1:6.4) due to different exposure regions as well as the different densities of population and wealth across the exposure maps.

Further, the years 2015 and 2014 recorded the smallest insured losses. The sum of the loss in both provinces did not exceed $62 and $140 million, respectively. Interestingly, QC did not experience a wind catastrophe in 2015 (Fig. 3).

Table 1. Top 5 most devastating wind catastrophes in the period 2010–2020 sorted according to the sum of the imposed insured loss in ON and QC.

| Rank | Year | Event Type | Location | Insured Loss [both provinces combined] (Million $) |
|------|------|------------|----------|---------------------------------------------------|
| 1 | 2018 | Convective storm | Hamilton, Toronto and the GTA | 647.17 |
| 2 | 2016 | Convective storm | Windsor | 372.02 |
| 3 | 2018 | Tornado | Dunrobin, ON; Ottawa, ON, Nepean, ON; Gatineau, QC | 340.77 |
| 4 | 2018 | Winter storm | The GTA, Leamington, Hamilton, Kitchener, Guelph, London, Chatham-Kent, Ottawa Gatineau, Waterloo | 242.30 |
| 5 | 2019 | Convective storm | Niagara Region, eastern Ontario, and areas surrounding Montreal | 242.15 |

Linear regressions shown in Fig. 3 demonstrate that the number of events that occurred in QC in the period 2010–2020 has an increasing trend with small fluctuations around the trendline, resulting in the standard deviation of 1.33. Similarly, the imposed losses in QC also show a positive trend with a standard deviation of $65 million. The ON data, however, reveals an interesting trend: While the number of catastrophes indicates a downward trend in the recent years compared to the first half of the decade, the insured losses are characterized by an increasing trend. Since the raw data were normalized to adjust for the wealth accumulation over time (Section 2), the obtained upward trend of wind damage and the high standard deviation of $317 million in ON can be interpreted in terms of the increased intensity of wind events as well as the increase in the scale of their impact. The linear regressions in Fig. 5 provide additional insights on the interaction of insured loss and the number of wind events. This analysis shows that the average loss per wind catastrophe has been increasing over time in both provinces. Figures like this emphasize the importance of climate, wind engineering, and atmospheric boundary layer research focusing on urban sustainability and resilience. However, it should also be noted that the Mann-Kendall test of trend significance (Romanic et al., 2015) performed at the significance level of $\alpha = 0.1$ shows that only the trend of average loss per event for ON is statistically significant.

## 3.3 Seasonal analysis of wind catastrophes

Fig. 6 depicts the number of events and wind damage recorded in different seasons. The spring seasons in ON were the most critical period during 2010–2020, when almost 36% and 38% of the total number of catastrophic wind events and total insured loss occurred, respectively. On the other hand, while QC experienced the highest number of wind catastrophes (=8) in this season, the maximum damage occurred in the fall seasons. Once again, we observe that the discrepancy in the number of events (Fig. 6a) does not directly translate to the inflicted wind damage (Fig. 6b), particularly in the spring season when the activity of wind storms is the highest. Interestingly, the number of occurred catastrophes in the fall and winter seasons is the same between the provinces. Moreover, winter seasons have been calm periods in ON and QC. The number of wind catastrophes in this season is 0.30 (ON) and 0.43 (QC) of the average number of occurred events in the other three seasons. Moreover, the damage across ON and QC in winter is respectively 0.43 and 0.19 of the average loss in other seasons. These results demonstrate that the activity of wind catastrophes in winter is substantially lower than in the rest of the year.

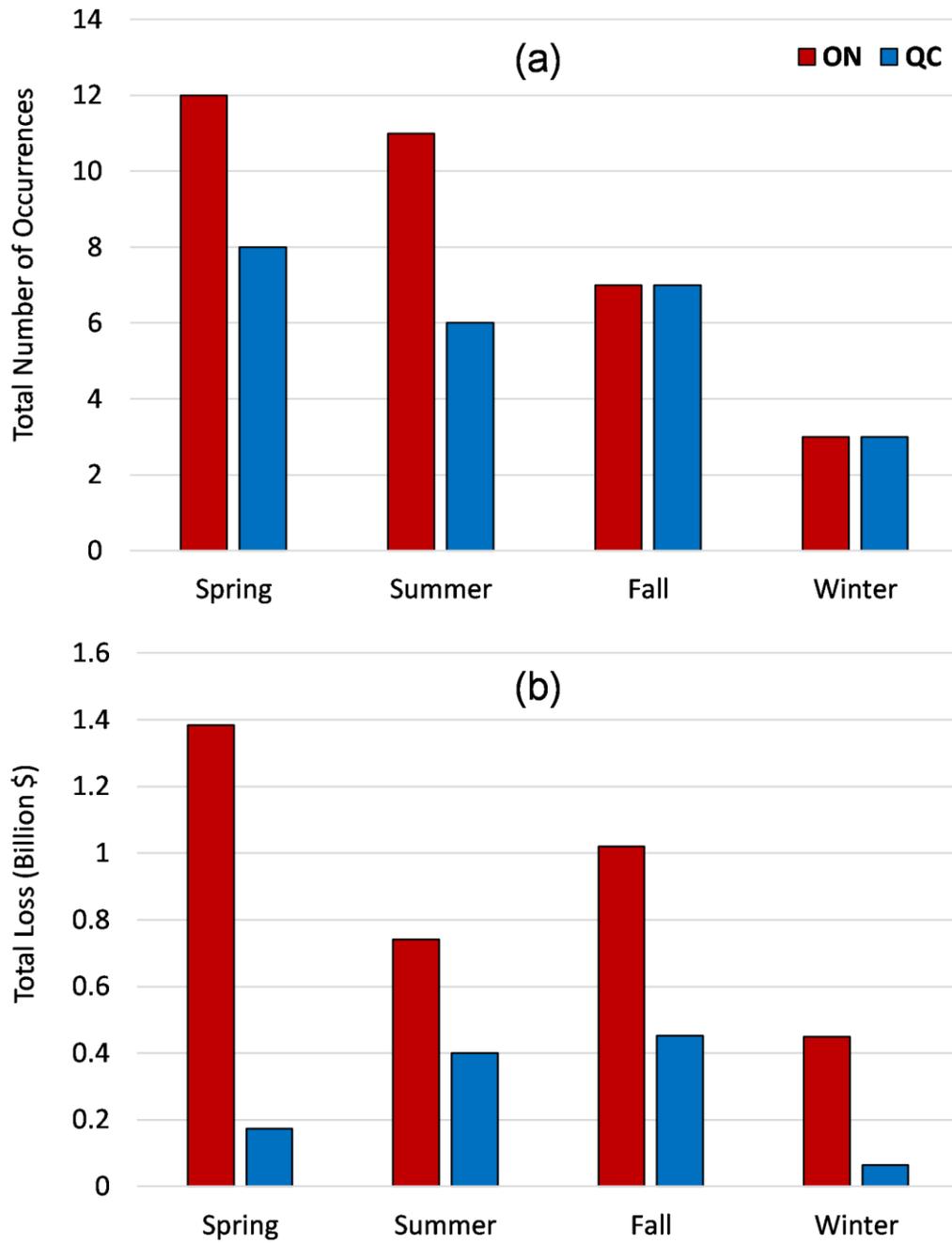

Fig. 6. Seasonal analysis of wind catastrophes for ON and QC in the period 2010–2020 based on (a) the total number of events and (b) total wind damage.

### 3.4 Damage classification by windstorm type

Fig. 7 compares the occurrence and loss data of four investigated wind types that caused the reported wind catastrophes. Convective storms stand as the most devastating wind type by inflicting more than $2.36 and $0.79 billion in damage in ON and QC between 2010–2020. The two provinces received 22 (ON) and 14 (QC) catastrophic convective wind events in the investigated period. Convective storms cost ON three times more than QC, mainly due to their eastward and northeastward movement (Fig. 4) that resulted in higher severity of storms over ON. As shown in Fig. 8, convective storms comprised nearly three-quarters of the overall loss in ON and QC by accounting for more than half of the total number of catastrophic wind events in these provinces. Moreover, Fig. 9 reveals that at least one convective storm occurred each year in 2010-2020 (except for 2010 and 2015 for QC). On average, ON and QC were impacted by 2 and 1.3 convective storms per year in the studied period, causing more than $214 and $71 million in damage per year, respectively.

To explore further the identified significance of convective storms in ON and QC, Fig. 10 focuses on their trends during the period 2010–2020. We demonstrate that the insured losses imposed by convective storms have increased over time, although the annual number of these events has experienced a slight downward trend in ON. Interestingly, this downward trend of violent convective storms contradicts the popular narrative that the changing climate will cause more severe storms (Estrada et al., 2015; Grinsted et al., 2019; Nordhaus, 2010; Sander et al., 2013). However, according to World Meteorological Organization (WMO, 2011, 2016) the length of climate study dataset should be at least 30 years, we recognize that more than 20 years of data is needed to additionally test this hypothesis. Further inspection of Fig. 3 and Fig. 10 highlights that the trend of convective storms mainly shaped the annual number of wind events and the associated total wind damage. From the wind engineering point of view, our analysis demonstrates that the high-end losses in ON and QC are governed by convective windstorms. This finding is somewhat different from the current focus of wind loading standards and recommendations (ASCE/SEI 7-10, 2013; NBCC, 2015) that primarily target synoptic winds. Fig. 7 shows that the catastrophic damage caused by large-scale atmospheric boundary layer winds is substantially smaller than the damage inflicted by non-synoptic winds, such as convective storms and tornadoes. Indeed, codification of thunderstorm winds and tornadoes is substantially more challenging than that of synoptic winds due to the highly localized, transient and non-Gaussian nature of thunderstorm winds and tornadoes (Hangan et al., 2019).

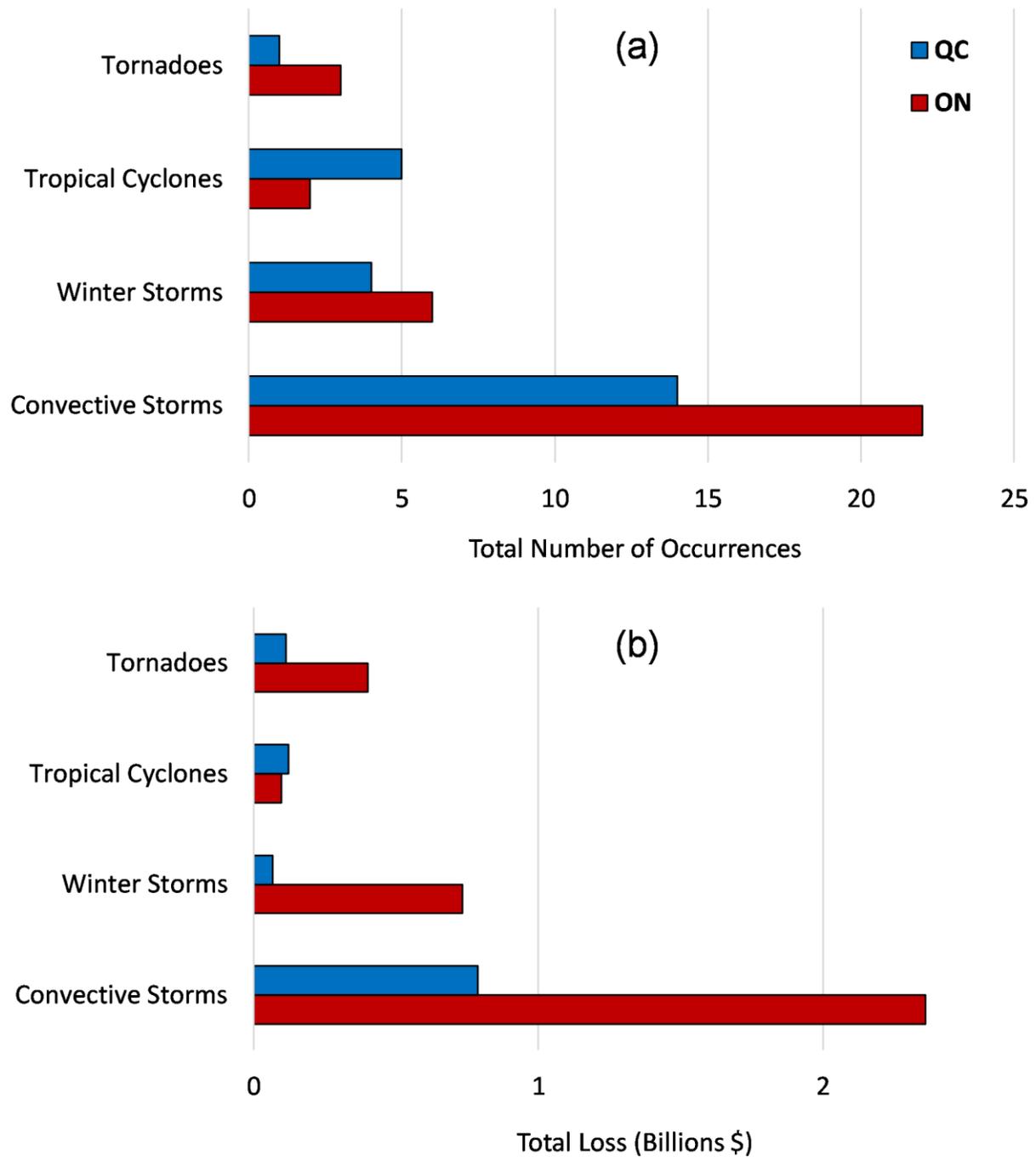

Fig. 7. Comparison of four investigated wind types causing wind catastrophes in ON and QC based on the (a) total number of events and (b) total inflicted damage.

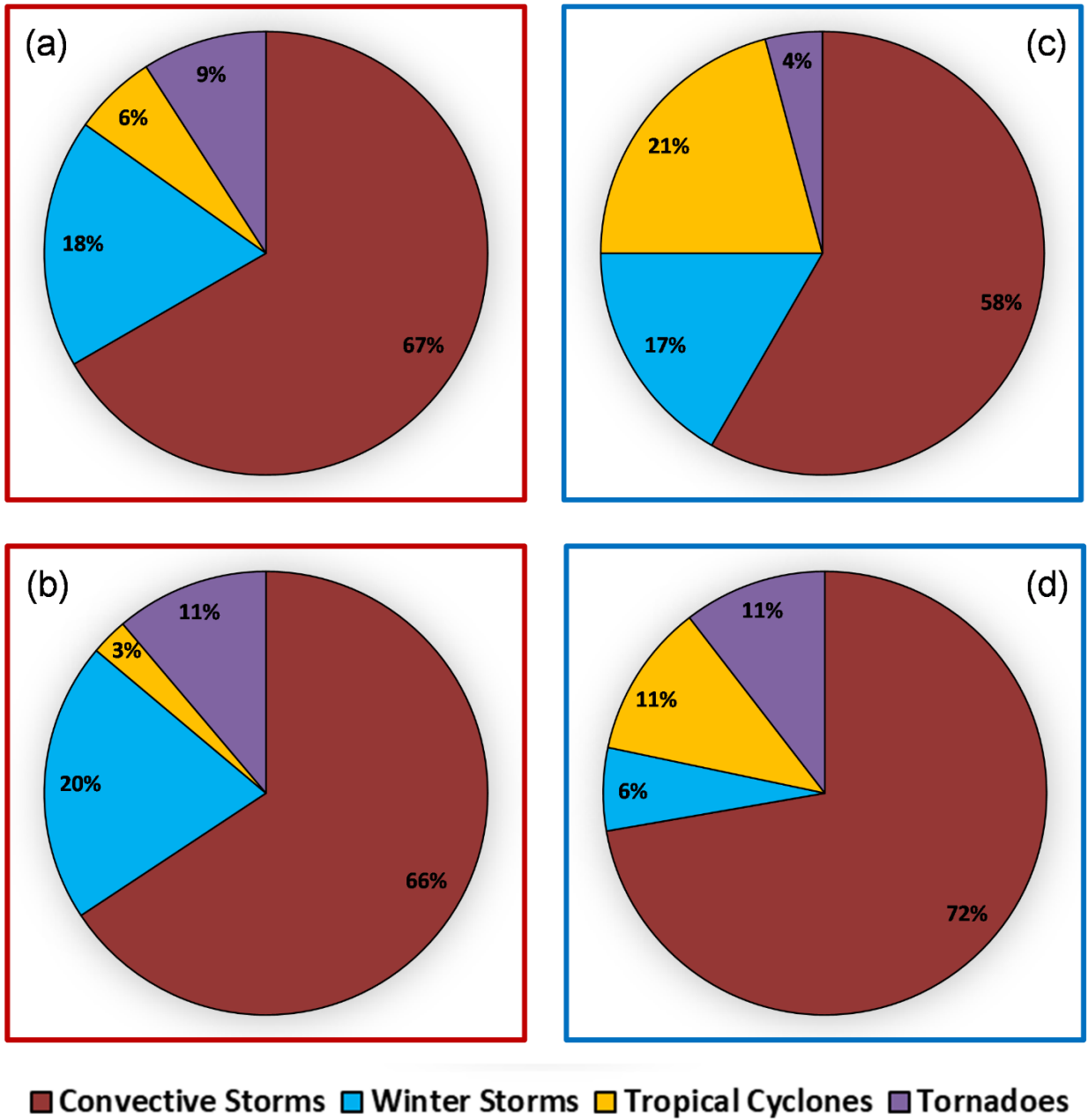

Fig. 8. Comparison of four investigated wind types causing wind catastrophes based on their contribution in (a) total number of events and (b) inflicted damage for ON. The panels (c) and (d) are the same as (a) and (b) but for QC.

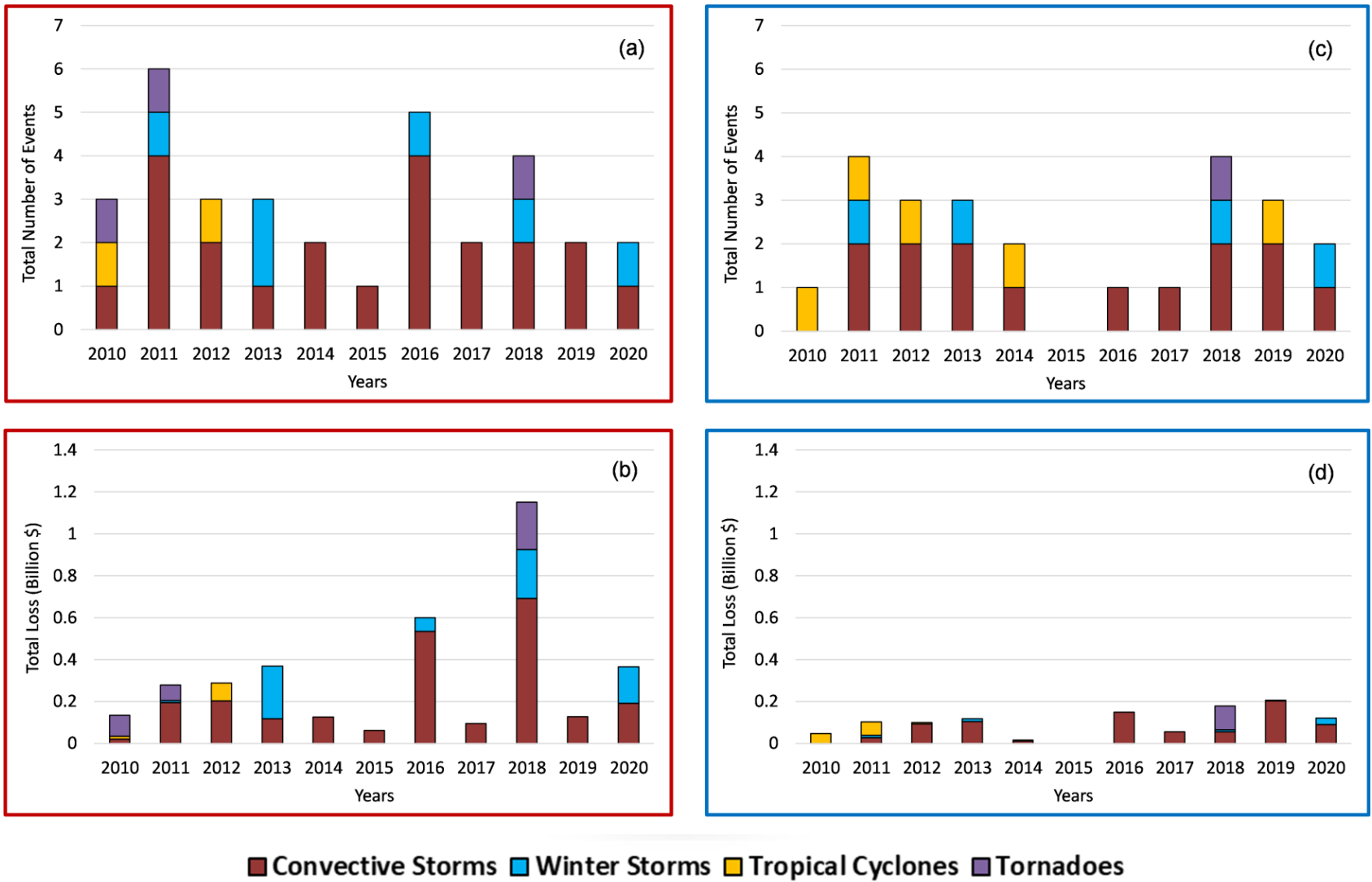

Fig. 9. Yearly comparison of four investigated wind types causing wind catastrophes in the period 2010–2020 based on the (a) total number of events and (b) total damage for ON. The panels (c) and (d) are the same as (a) and (b) but for QC.

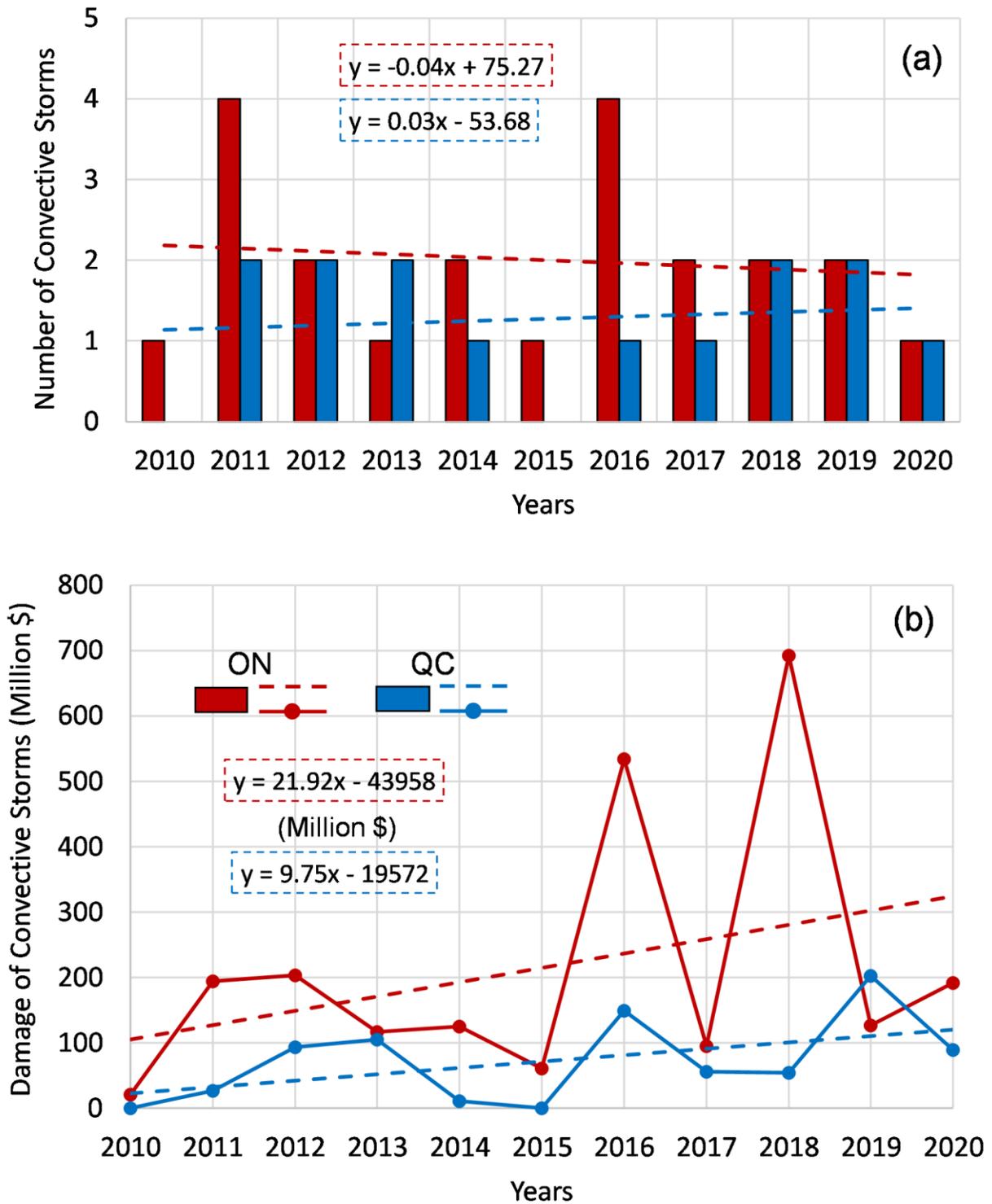

Fig. 10. Yearly trends of (a) total number of convective storms and (b) total inflicted damage by them for ON and QC in the period 2010–2020. The straight lines are linear trends with the trend equations provided in the plot.

Fig. 8 shows that winter storms were responsible for one-fifth of the total damage, ranking them in second place of the most destructive wind events in ON and QC. This high ranking of winter storms is mainly due to several major winter storms that occurred in 2013 (2 events) and 2018 (1 event) (see Table 1 and Fig. 9) which together made up for 66% of the total winter storm losses in the period 2010–2020 in ON. On the other hand, winter storms were the least disastrous wind events in QC (Fig. 7), and the inflicted losses did not exceed $66 million. Notice in Fig. 4 that all winter storms had eastward or northeastward trajectories that led to more severe damage in ON.

Tropical cyclones had the least impact on the total damage in ON. Fig. 4 shows that tropical cyclones mainly affected the eastern parts of QC, and they moved mostly northward. Hence, they were responsible for the smallest loss in ON among all other wind types. Hurricane Nicole in 2010 and hurricane Sandy in 2012 were the only catastrophic tropical cyclones to have an impact in ON in the period 2010–2020 (Fig. 9). In both instances, they only affected a small part of ON thereby causing the total loss value below $100 million (Fig. 7). In addition, while the damage data for QC show a higher frequency of catastrophic hurricanes than in ON, the total losses were fairly similar between the provinces (Fig. 7, around $122 million per province).

Fig. 9 shows that QC and ON experienced only one and three catastrophic tornadoes in 2010–2020, respectively. These wind events were responsible for almost one-tenth of the total loss in each province (Fig. 8). As depicted in Fig. 9, imposed damage by tornadoes was most notable in 2018, when Dunrobin, Ottawa, and Nepean regions in ON, as well as the Gatineau region in QC, were affected by large tornado outbreak that spawned six tornadoes in one day. The strongest tornado was an EF3-rated twister that moved through the suburban Ottawa neighborhoods of Kinburn and Dunrobin (ON) before moving on to Gatineau (QC). This event is ranked as the third most detrimental catastrophe in the period 2010–2020 by inflicting more than $340 million in damage in ON and QC combined (Table 1).

Fig. 11 shows the average loss per event for all investigated wind types. The maximum average loss per catastrophe in both provinces was caused by tornadoes. This analysis shows that while tornados were not as frequent as some other catastrophic wind events (e.g., convective storms) (Fig. 8), each tornadic event was extremely destructive and caused more than $133 and $114 million in damage in ON and QC, respectively. As demonstrated in Fig. 12, tornados were characterized by the highest gust speeds reported among all wind-related catastrophes. Based on Environment & Climate Change Canada (ECCC) storm summary reports, there were only four instances of wind gusts exceeding 70 m s$^{-1}$, but three of these four events were tornadoes that struck Leamington in 2010, Goderich in 2011, and Ottawa and Gatineau in 2018. An example of damage caused by the Goderich tornado is shown Fig. 1(d).

Tornadoes also caused the most similar value of losses among the two provinces. On the other hand, the highest discrepancy in the average loss per event is caused by winter storms. In QC, winter storms caused approximately $16.5 million in damage per event, resulting in the smallest average loss among all wind types. However, ON experienced six winter storm-related catastrophes over the same period (Fig. 9). Among these, the 2018 winter storm was extremely damaging (Table 1) and inflicted more than $232 million in loss.

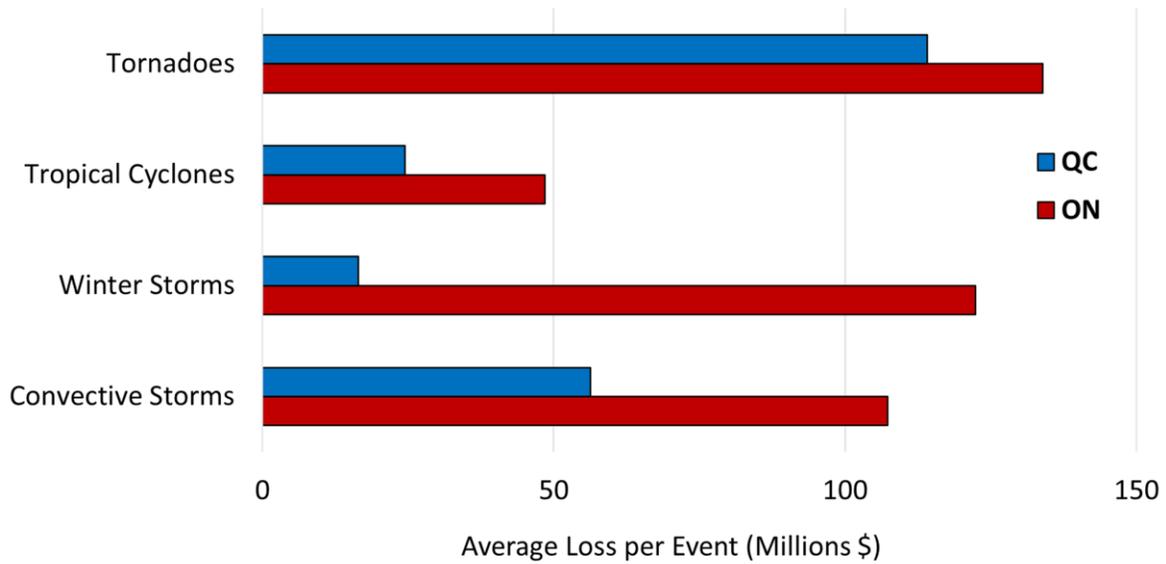

Fig. 11. Average wind damage per event caused by the four investigated wind types.

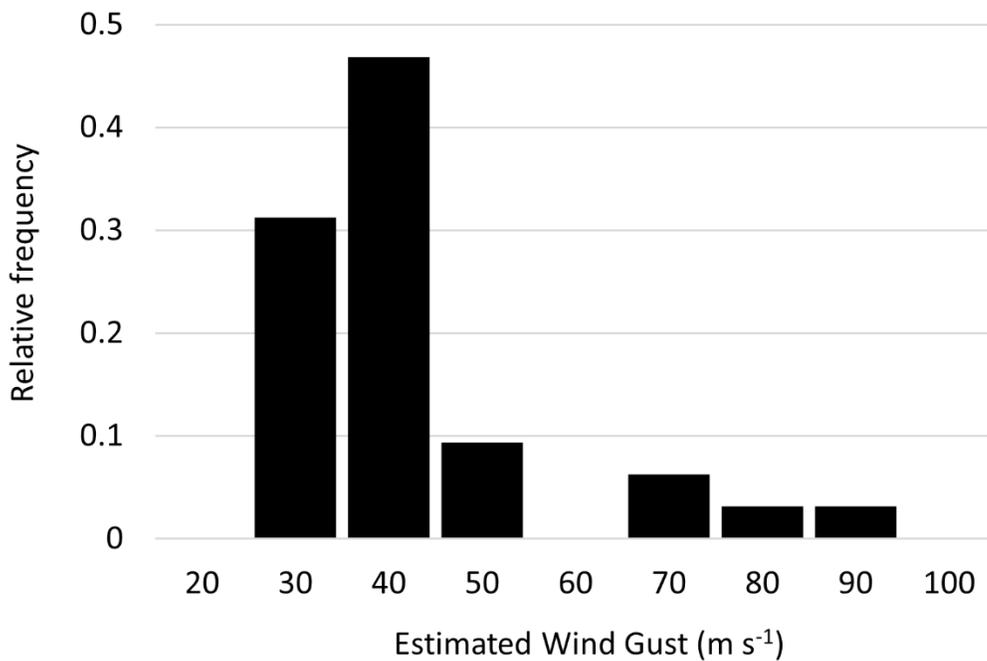

Fig. 12. The histogram of highest wind gusts associated with the analyzed catastrophic wind events in ON and QC in the period 2010–2020. The values come from the ECCC storm summary reports.

It should be remarked that physical damage comprised more than 97% of total losses in most cases (based on the available data for the period 2012–2020). Exceptionally, winter storms and tornadoes that occurred in QC had a higher ratio of non-physical damage, which were about 8.5% and 12%, respectively. Physical damage is loss to a structure or object while non-physical damage represents additional living expenses for individuals or business interruption for commercial property.

Fig. 13 shows the seasonal comparison of wind types based on the number of events and their associated loss. Convective storms were responsible for the most significant portion of the total damage and the highest number of events throughout a year, except in the winter seasons when the winter storms were the dominant wind type in both provinces. Some winter storms also occurred in early spring (mostly March) when the weather conditions are still supportive for the development of snow and winter storms. On the other hand, tornados occurred mainly in summer and early fall (September), when the atmospheric conditions are most favorable for the development of deep convection including supercells as well as mesoscale convective systems. Finally, tropical cyclones affected ON and QC during the hurricane season in late summer and early fall.

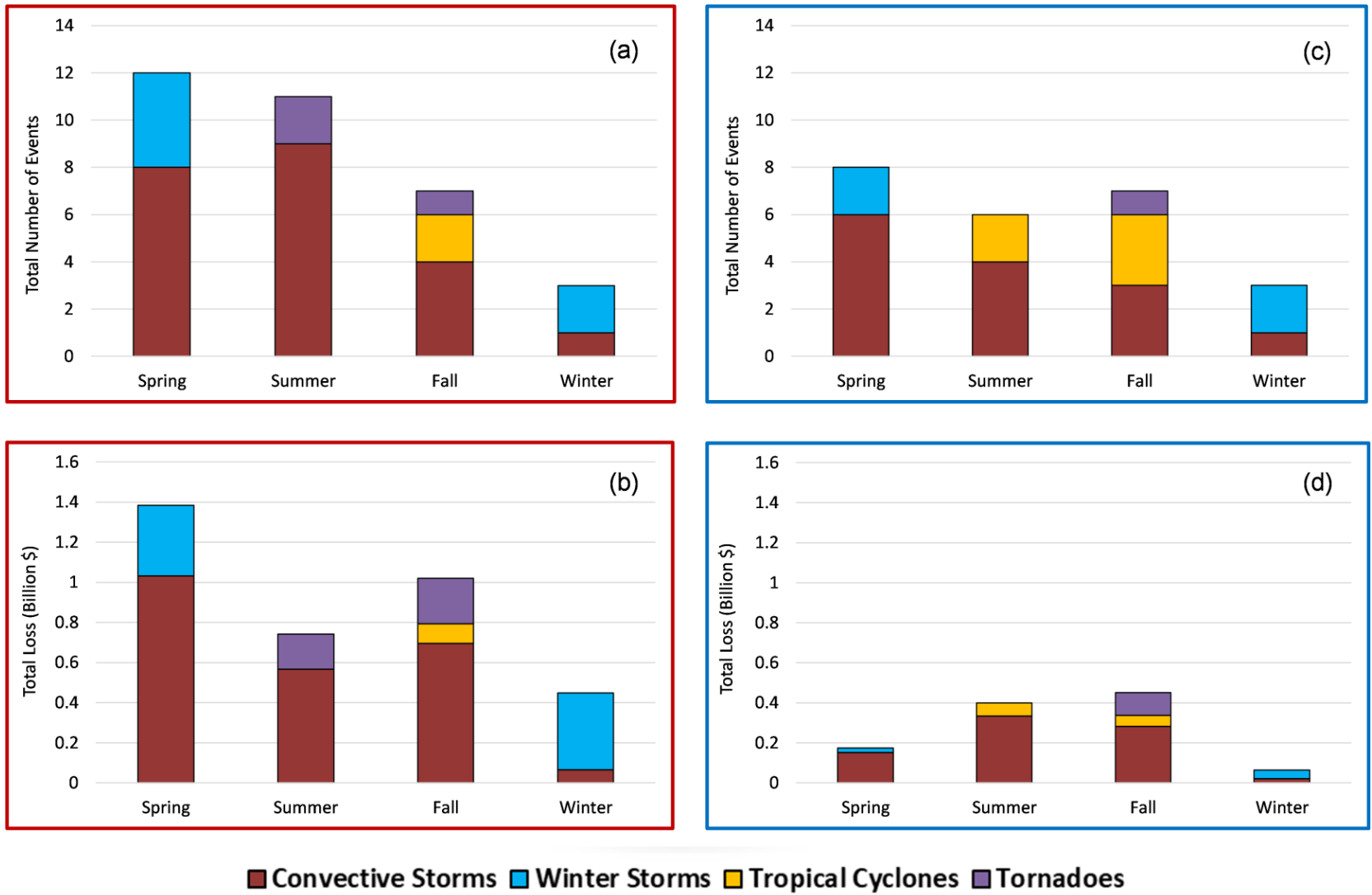

Fig. 13. Seasonal comparison of four investigated wind types causing wind catastrophes based on the (a) total number of events and (b) total inflicted loss for Ontario; (c) and (d) same as (a) and (b) but for Quebec.

### 3.5 Damage type classification: personal, commercial, and auto

Fig. 14 classifies the wind damage into three categories: (1) personal; (2) commercial; and (3) auto. Personal damage is physical or non-physical damage to personal property. For example, personal property can include a home, cottage, outbuilding (i.e., shed). Commercial damage is physical or non-physical damage to commercial property. For example, commercial property can include a commercial building, farm, and other commercial structures and objects. The auto category represents the sum of personal and commercial automobile losses related to wind catastrophes, excluding accidents that were not related to wind catastrophes. Considering both provinces combined, personal damage contributed nearly 70% of total all losses, while the commercial and auto losses respectively accounted for about 18% and 12% of the total.

In addition, Fig. 15 shows a similar classification for each of the four considered wind types. Personal damage comprised the most significant part of inflicted losses in most cases. For example, personal damage reached nearly 90% of the overall loss from all tropical cyclones in QC. Commercial damage did not exceed one-fifth of the total loss in all cases; excluding tornadoes, in which case the total losses were evenly distributed between personal and commercial lines of business when analyzing both provinces combined. Finally, auto losses accounted for less than one-tenth of total loss in most cases, except in the case of convective storms in QC when a quarter of losses was auto damage.

By noticing that ~97% of total losses were due to the physical damage to various structure and properties, this analysis highlights the negative effect of severe windstorms on people and communities. We believe that the present study has demonstrated the need for more research on resilience of communities to severe winds, particularly non-synoptic wind events caused by convective storms. Some of the topics that deserve more research are modelling of the interaction between severe winds caused by convective storms and urban environments as well as vulnerability of cities and urban blocks to severe non-synoptic winds.

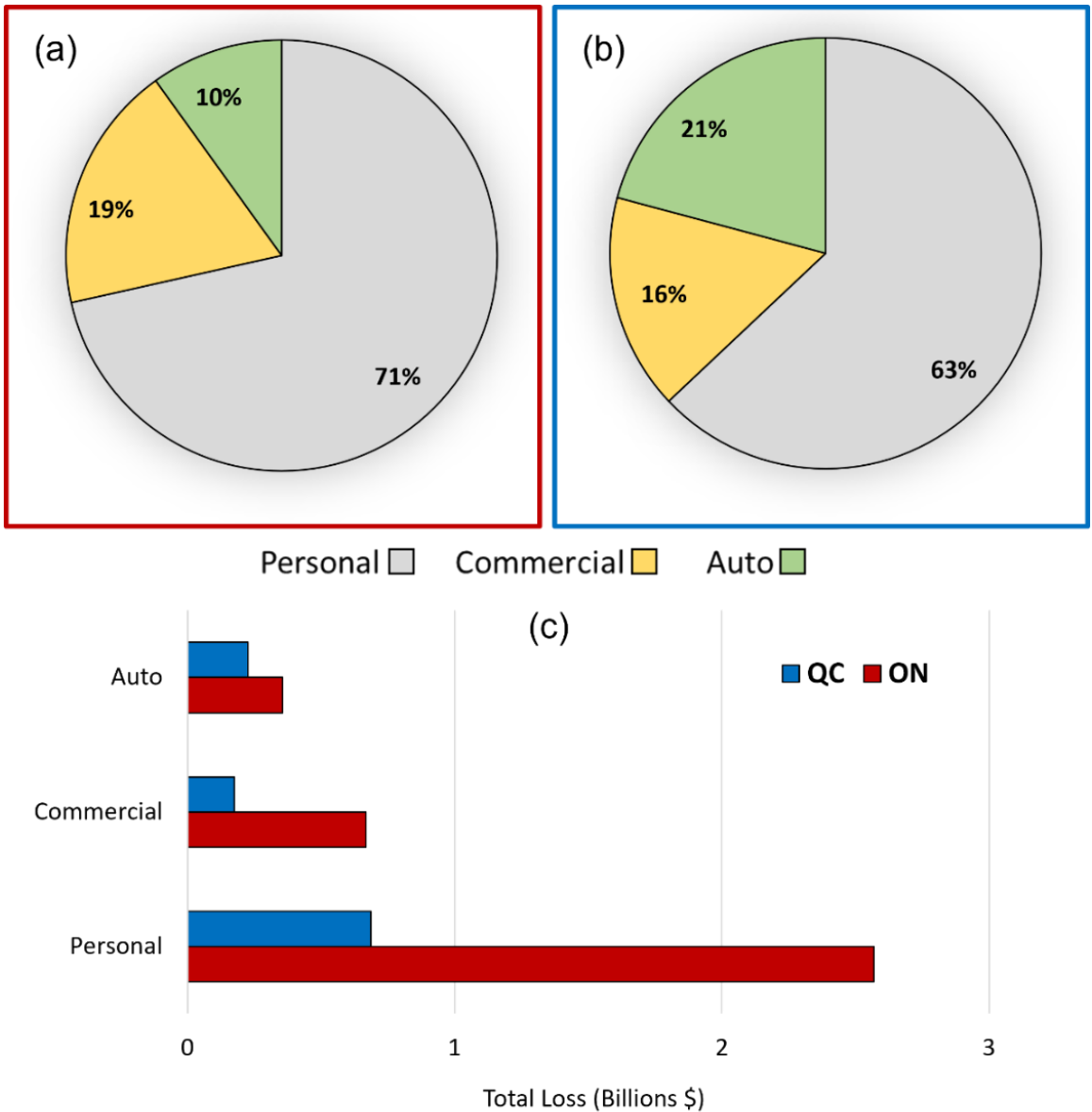

Fig. 14. Classification of wind damage in ON and QC in the period 2010–2020 based on the three categories of personal, commercial, and auto losses.

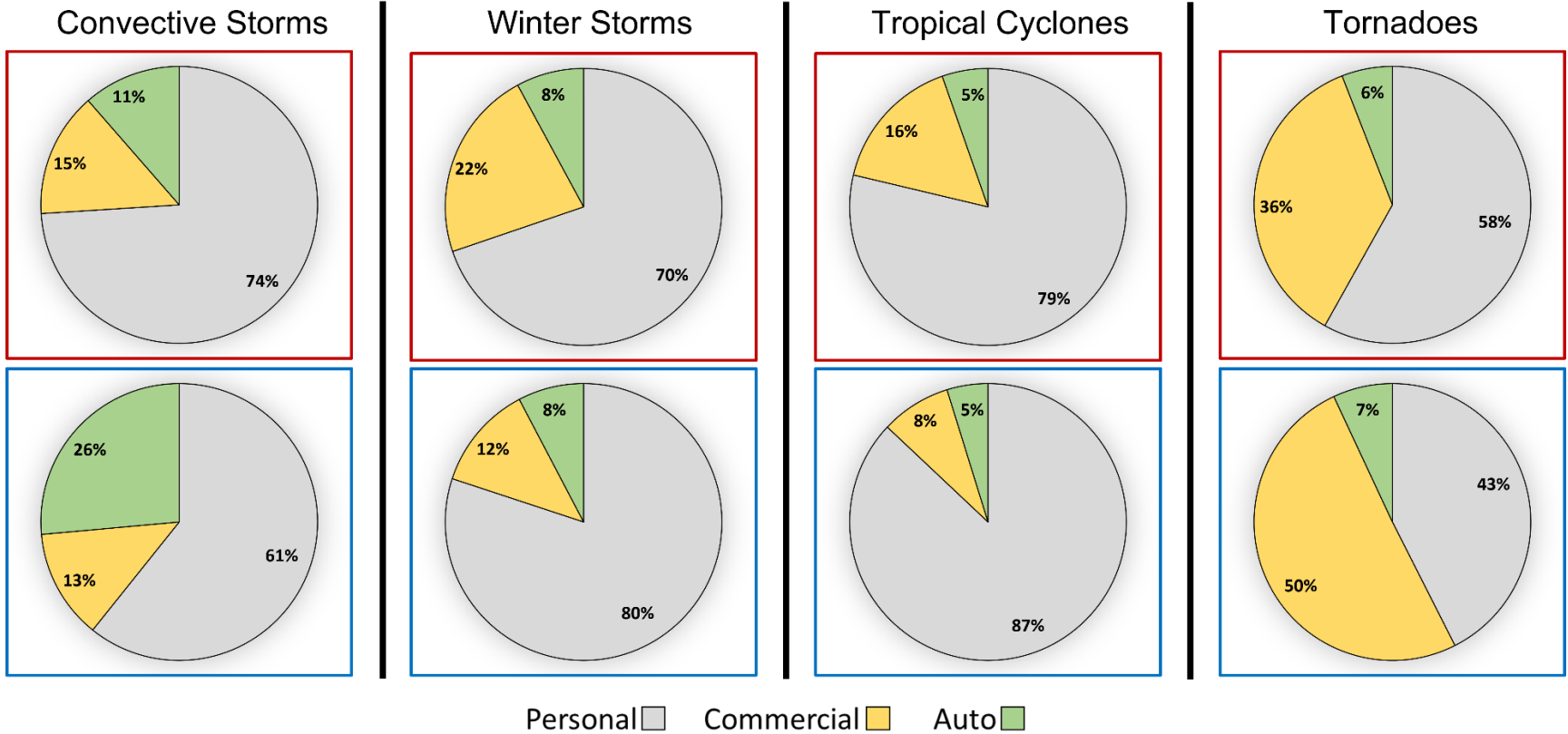

Fig. 15. Classification of wind damage in ON and QC in the period 2010–2020 based on the three categories of personal, commercial, and auto losses for each of the four investigated wind types.

## 4    Summary and conclusions

The goal of this study was to systematically analyze windstorm-induced catastrophic damage during the period 2010–2020 and for two Canadian provinces of Ontario (ON) and Quebec (QC). Here, we followed the insurance market definitions of catastrophic events and defined them as those wind events that caused losses of more than CA$25 million for the insurance industry (hereafter we omit CA because all losses are reported in Canadian dollars).

Our study was based on the estimated insurance loss data obtained from the CatIQ database. The reported losses were first normalized to the reference year 2019 in order to compensate for the increased population and asset values over time. This study identified and analyzed a total of 39 catastrophic wind events over ON and QC in the period 2010–2020. These events also accounted for nearly three-quarters of all catastrophes in ON and QC in the investigated period. In addition, the total imposed wind losses were more than $4.5 billion, encompassing over 64% of total losses induced by all catastrophic perils in ON and QC between 2010 and 2020. Almost 97% of the total damage was physical. Moreover, personal damage dominated total losses by standing for near 70% of wind damage, while commercial and auto damage portions did not exceed 20% and 10%, respectively.

The distribution of losses between the two provinces can be summarized through the following QC:ON ratios:

- 1:3.3—total wind damage losses induced by all events;
- 1:1.4—total number of catastrophic events;
- 1:2.4—average loss per catastrophic event.

Clearly, ON sustained more wind damage than QC. The peak of insured losses occurred in 2018 when ON experienced large damage of about $1.15 billion. On the other hand, 2015 and 2014 recorded the smallest losses, and QC did not experience a single wind catastrophe in 2015. The seasonal analysis showed that the spring season was the most critical period that accounted for almost 40% of total annual losses. The activity of windstorms was the lowest in winter. The highest-ranked wind catastrophe in the studied period was a 2018 severe convective storm that caused approximately $650 million in insured loss, mostly due to the damage concentrated around the Great Toronto Area and Hamilton in ON.

In terms of different windstorm types, the total losses caused by the convective storms, winter storms, tornadoes, and tropical cyclones in ON and QC are respectively:

- ON: 66%, 20%, 11%, 3%; and
- QC: 72%, 6%, 11%, 11%.

Therefore, the wind catastrophes in ON and QC have been governed by the convective storms, inducing more than $3 billion in damage by accounting for over half of the total number of catastrophes in these two provinces.

Lastly, we showed that tornadoes caused the highest average loss per catastrophe in both provinces. While tornadic events are relatively infrequent when compared to other wind perils, each tornadic event was extremely destructive, inflicting almost $133 and $114 million in damage in ON and QC, respectively.

Acknowledgments

The authors thank the CatIQ personnel, in particular Ms. Laura Twidle, for their assistance in retrieving the relevant data from the CatIQ platform. We acknowledge the support of the Natural Sciences and Engineering Research Council of Canada (NSERC), Discovery Grant RGPIN-2021-02651 and the Wares Innovation Prospectors Fund (2020).

Declaration of Competing Interest

The authors declare that they have no known competing financial interests or personal relationships that could have appeared to influence the work reported in this paper.

References

ASCE/SEI 7-10. (2013). Minimum Design Loads for Buildings and Other Structures (ASCE/SEI 7-10). *American Society of Civil Engineers*. https://doi.org/10.1061/9780784412916

Barredo, J. (2010). No upward trend in normalised windstorm losses in Europe. *Nat Hazards Earth Syst Sci*, *10*, 97-104. https://doi.org/10.5194/nhess-10-97-2010

Barthel, F., & Neumayer, E. (2012). A trend analysis of normalized insured damage from natural disasters. *Climatic Change*, *113*(2), 215-237. https://doi.org/10.1007/s10584-011-0331-2

Benjamin, G. (2019). A crane has collapsed at the corner of South Park and Spring Garden. #Halifax. In *Twitter (@GlobalGraeme)*. https://twitter.com/GlobalGraeme/status/1170424418260180992?ref_src=twsrc%5Etfw%7Ctwcamp%5Etweetembed%7Ctwterm%5E1170424418260180992%7Ctwgr%5E%7Ctwcon%5Es1_&ref_url=https%3A%2F%2Fglobalnews.ca%2Fnews%2F5872891%2Fcanadian-forces-post-dorian%2F.

Boruff, B. J., Easoz, J. A., Jones, S. D., Landry, H. R., Mitchem, J. D., & Cutter, S. L. (2003). Tornado hazards in the United States. *Climate Research*, *24*(2), 103-117. https://doi.org/10.3354/cr024103

Bouwer, L. M., & Wouter Botzen, W. (2011). How sensitive are US hurricane damages to climate? Comment on a paper by WD Nordhaus. *Climate Change Economics*, *2*(01), 1-7. https://doi.org/10.1142/S2010007811000188

Brooks, H. E., & Doswell III, C. A. (2001). Normalized damage from major tornadoes in the United States: 1890–1999. *Weather and Forecasting*, *16*(1), 168-176. https://doi.org/10.1175/1520-0434(2001)016<0168:NDFMTI>2.0.CO;2

Burlando, M., Romanic, D., Boni, G., Lagasio, M., & Parodi, A. (2020). Investigation of the Weather Conditions During the Collapse of the Morandi Bridge in Genoa on 14 August 2018 Using Field Observations and WRF Model. *Atmosphere*, *11*(7), 724. https://doi.org/10.3390/atmos11070724

Cardona, O.-D., Ordaz, M. G., Mora, M. G., Salgado-Gálvez, M. A., Bernal, G. A., Zuloaga-Romero, D., Fraume, M. C. M., Yamín, L., & González, D. (2014). Global risk assessment: A fully probabilistic seismic and tropical cyclone wind risk assessment. *International journal of disaster risk reduction*, *10*, 461-476. https://doi.org/10.1016/j.ijdrr.2014.05.006

CatIQ. (2021). *Catastrophe Indices and Quantification Inc.* Canada's Loss And Exposure Indices Provider. . https://public.catiq.com/

CCCR. (2019). Canada's Changing Climate Report.

Changnon, S. A. (2001). Damaging thunderstorm activity in the United States. *Bulletin of the American Meteorological Society*, *82*(4), 597-608. https://doi.org/10.1175/1520-0477(2001)082<0597:DTAITU>2.3.CO;2

Chen, W., Lu, Y., Sun, S., Duan, Y., & Leckebusch, G. C. (2018). Hazard footprint-based normalization of economic losses from tropical cyclones in china during 1983–2015. *International Journal of Disaster Risk Science*, *9*(2), 195-206. https://doi.org/10.1007/s13753-018-0172-y

Dai, K., Chen, S.-E., Luo, M., & Loflin Jr, G. (2017). A framework for holistic designs of power line systems based on lessons learned from Super Typhoon Haiyan. *Sustainable cities and society*, *35*, 350-364. https://doi.org/10.1016/j.scs.2017.08.006

Davenport, A. G. (1961). The application of statistical concepts to the wind loading of structures. *Proceedings of the Institution of Civil Engineers*, *19*(4), 449-472. https://doi.org/10.1680/iicep.1961.11304

Diffenbaugh, N. S., Scherer, M., & Trapp, R. J. (2013). Robust increases in severe thunderstorm environments in response to greenhouse forcing. *Proceedings of the National Academy of Sciences*, *110*(41), 16361-16366. https://doi.org/10.1073/pnas.1307758110

Estrada, F., Botzen, W. W., & Tol, R. S. (2015). Economic losses from US hurricanes consistent with an influence from climate change. *Nature Geoscience*, *8*(11), 880-884. https://doi.org/10.1038/ngeo2560

Fischer, T., Su, B., & Wen, S. (2015). Spatio-temporal analysis of economic losses from tropical cyclones in affected provinces of China for the last 30 years (1984–2013). *Natural Hazards Review*, *16*(4), 04015010. https://doi.org/10.1061/(ASCE)NH.1527-6996.0000186

Global News. (2014). https://globalnews.ca/news/1398812/severe-thunderstorm-watch-issued-for-toronto-southern-ontario/. In.

Global News. (2018). https://globalnews.ca/news/4146552/toronto-power-outage-transportation-delays/. In.

Grinsted, A., Ditlevsen, P., & Christensen, J. H. (2019). Normalized US hurricane damage estimates using area of total destruction, 1900− 2018. *Proceedings of the National Academy of Sciences*, *116*(48), 23942-23946. https://doi.org/10.1073/pnas.1912277116

Gu, D., Zhao, P., Chen, W., Huang, Y., & Lu, X. (2021). Near real-time prediction of wind-induced tree damage at a city scale: Simulation framework and case study for Tsinghua University campus. *International journal of disaster risk reduction*, *53*, 102003. https://doi.org/10.1016/j.ijdrr.2020.102003

Gumaro, J. J. C., Acosta, T. J. S., Tan, L. R. E., Agar, J. C., Tingatinga, E. A. J., Musico, J. K. B., Plamenco, D. A. D., Ereño, M. N. C., Pacer, J. S., & Villalba, I. B. O. (2022). Identification of key components for developing building types for risk assessment against wind loadings: The case of Cebu Province, Philippines. *International journal of disaster risk reduction*, *67*, 102686. https://doi.org/10.1016/j.ijdrr.2021.102686

Hangan, H., Romanic, D., & Jubayer, C. (2019). Three-dimensional, non-stationary and non-Gaussian (3D-NS-NG) wind fields and their implications to wind–structure interaction problems. *Journal of Fluids and Structures*, *91*, 102583. https://doi.org/10.1016/j.jfluidstructs.2019.01.024

Harrison, S., Silver, A., & Doberstein, B. (2015). Post-storm damage surveys of tornado hazards in Canada: Implications for mitigation and policy. *International journal of disaster risk reduction*, *13*, 427-440. https://doi.org/10.1016/j.ijdrr.2015.08.005

Henstra, D. (2012). Toward the climate-resilient city: extreme weather and urban climate adaptation policies in two Canadian provinces. *Journal of Comparative Policy Analysis: Research and Practice*, *14*(2), 175-194. https://doi.org/10.1080/13876988.2012.665215

Huang, Q., Jiang, W., & Hong, H. (2021). Statistical Assessment of Spatial Tornado Occurrences in Canada: Modeling and Estimation. *Journal of Applied Meteorology and Climatology*, *60*(12), 1633-1651. https://doi.org/10.1175/JAMC-D-20-0141.1

Hunt, J. (2004). How can cities mitigate and adapt to climate change? *Building Research & Information*, *32*(1), 55-57. https://doi.org/10.1080/0961321032000150449

Kelly, D. L., Schaefer, J. T., & Doswell III, C. A. (1985). Climatology of nontornadic severe thunderstorm events in the United States. *Monthly weather review*, *113*(11), 1997-2014. https://doi.org/10.1175/1520-0493(1985)113<1997:CONSTE>2.0.CO;2

Klotzbach, P. J., Bowen, S. G., Pielke, R., & Bell, M. (2018). Continental US hurricane landfall frequency and associated damage: Observations and future risks. *Bulletin of the American Meteorological Society*, *99*(7), 1359-1376. https://doi.org/10.1175/BAMS-D-17-0184.1

Martinez, A. B. (2020). Improving normalized hurricane damages. *Nature Sustainability*, *3*(7), 517-518. https://doi.org/10.1038/s41893-020-0550-5

McBean, G. (2004). Climate change and extreme weather: a basis for action. *Natural Hazards*, *31*(1), 177-190. https://doi.org/10.1023/B:NHAZ.0000020259.58716.0d

Miller, S., Muir-Wood, R., & Boissonnade, A. (2008). An exploration of trends in normalized weather-related catastrophe losses. *Climate extremes and society*, *12*, 225-247. https://doi.org/10.1017/CBO9780511535840.015

Mohleji, S., & Pielke Jr, R. (2014). Reconciliation of trends in global and regional economic losses from weather events: 1980–2008. *Natural Hazards Review*, *15*(4), 04014009. https://doi.org/10.1061/(ASCE)NH.1527-6996.0000141

NBCC. (2015). *National building code of Canada 2015*. National Research Council Canada.

Neumayer, E., & Barthel, F. (2011). Normalizing economic loss from natural disasters: A global analysis. *Global Environmental Change*, *21*(1), 13-24. https://doi.org/10.1016/j.gloenvcha.2010.10.004

Nordhaus, W. D. (2010). The economics of hurricanes and implications of global warming. *Climate Change Economics*, *1*(01), 1-20. https://doi.org/10.1142/S2010007810000054

OECD. (2016). Global Insurance Market Trends. In.

OECD. (2020). *Insurance indicators: Penetration*. https://stats.oecd.org/Index.aspx?QueryId=25444

Pant, S., & Cha, E. J. (2019a). Potential changes in hurricane risk profile across the United States coastal regions under climate change scenarios. *Structural Safety*, *80*, 56-65. https://doi.org/10.1016/j.strusafe.2019.05.003

Pant, S., & Cha, E. J. (2019b). Wind and rainfall loss assessment for residential buildings under climate-dependent hurricane scenarios. *Structure and Infrastructure Engineering*, *15*(6), 771-782. https://doi.org/10.1080/15732479.2019.1572199

Pielke Jr, R. A., Gratz, J., Landsea, C. W., Collins, D., Saunders, M. A., & Musulin, R. (2008). Normalized hurricane damage in the United States: 1900–2005. *Natural Hazards Review*, *9*(1), 29-42. https://doi.org/10.1061/(ASCE)1527-6988(2008)9:1(29)

Pielke Jr, R. A., Rubiera, J., Landsea, C., Fernández, M. L., & Klein, R. (2003). Hurricane vulnerability in Latin America and the Caribbean: Normalized damage and loss potentials. *Natural Hazards Review*, *4*(3), 101-114. https://doi.org/10.1061/(ASCE)1527-6988(2003)4:3(101)

Pielke, R. (2019). Tracking progress on the economic costs of disasters under the indicators of the sustainable development goals. *Environmental Hazards*, *18*(1), 1-6. https://doi.org/10.1080/17477891.2018.1540343

Pielke, R. (2021). Economic 'normalisation' of disaster losses 1998–2020: a literature review and assessment. *Environmental Hazards*, *20*(2), 93-111. https://doi.org/10.1080/17477891.2020.1800440

Pielke, R., Landsea, C. W., Musulin, R. T., & Downton, M. (1999). Evaluation of catastrophe models using a normalized historical record: Why it is needed and how to do it. *Journal of Insurance Regulation*, *18*(2), 177-194.

Pielke, R. A., & Landsea, C. W. (1998). Normalized hurricane damages in the United States: 1925–95. *Weather and Forecasting*, *13*(3), 621-631. https://doi.org/10.1175/1520-0434(1998)013<0621:NHDITU>2.0.CO;2

Raghavan, S., & Rajesh, S. (2003). Trends in tropical cyclone impact: A study in Andhra Pradesh, India: A study in Andhra Pradesh, India. *Bulletin of the American Meteorological Society*, *84*(5), 635-644. https://doi.org/10.1175/BAMS-84-5-635

Rawlings, K. (2011). http://www.ontariostorms.com/showpost.php?p=3751&postcount=26. In.

Refan, M., Romanic, D., Parvu, D., & Michel, G. (2020). Tornado loss model of Oklahoma and Kansas, United States, based on the historical tornado data and Monte Carlo simulation. *International journal of disaster risk reduction*, *43*, 101369. https://doi.org/10.1016/j.ijdrr.2019.101369

Romanic, D., Ćurić, M., Jovičić, I., & Lompar, M. (2015). Long-term trends of the 'Koshava' wind during the period 1949–2010. *International Journal of Climatology*, *35*(2), 288-302. https://doi.org/10.1002/joc.3981

Romanic, D., Refan, M., Wu, C.-H., & Michel, G. (2016). Oklahoma tornado risk and variability: A statistical model. *International journal of disaster risk reduction*, *16*, 19-32. https://doi.org/10.1016/j.ijdrr.2016.01.011

Sander, J., Eichner, J., Faust, E., & Steuer, M. (2013). Rising variability in thunderstorm-related US losses as a reflection of changes in large-scale thunderstorm forcing. *Weather, Climate, and Society*, *5*(4), 317-331. https://doi.org/10.1175/WCAS-D-12-00023.1

Schmidt, S., Kemfert, C., & Höppe, P. (2009). Tropical cyclone losses in the USA and the impact of climate change—A trend analysis based on data from a new approach to adjusting storm losses. *Environmental Impact Assessment Review*, *29*(6), 359-369. https://doi.org/10.1016/j.eiar.2009.03.003

Seeley, J. T., & Romps, D. M. (2015). The effect of global warming on severe thunderstorms in the United States. *Journal of Climate*, *28*(6), 2443-2458. https://doi.org/10.1175/JCLI-D-14-00382.1

Simmons, K. M., Kovacs, P., & Kopp, G. A. (2015). Tornado damage mitigation: Benefit–cost analysis of enhanced building codes in Oklahoma. *Weather, Climate, and Society*, *7*(2), 169-178. https://doi.org/10.1175/WCAS-D-14-00032.1

Simmons, K. M., Sutter, D., & Pielke, R. (2013). Normalized tornado damage in the United States: 1950–2011. *Environmental Hazards*, *12*(2), 132-147. https://doi.org/10.1080/17477891.2012.738642

Solari, G., Repetto, M. P., Burlando, M., De Gaetano, P., Pizzo, M., Tizzi, M., & Parodi, M. (2012). The wind forecast for safety management of port areas. *Journal of Wind Engineering and Industrial Aerodynamics*, *104*, 266-277. https://doi.org/10.1016/j.jweia.2012.03.029

Statistics Canada. (2020). *Table 36-10-0222-01 Gross domestic product, expenditure-based, provincial and territorial, annual (x 1,000,000)*. https://doi.org/10.25318/3610022201-eng

Stucki, P., Brönnimann, S., Martius, O., Welker, C., Imhof, M., Von Wattenwyl, N., & Philipp, N. (2014). A catalog of high-impact windstorms in Switzerland since 1859. *Natural Hazards and Earth System Sciences*, *14*(11), 2867-2882. https://doi.org/10.5194/nhess-14-2867-2014

The World Bank. (2020). *Inflation, GDP deflator (annual %) - Canada*. https://data.worldbank.org/indicator/NY.GDP.DEFL.KD.ZG?end=2020&locations=CA&name_desc=false&start=1961&type=points&view=chart

United Nations. (2015). *The human cost of weather related disasters 1995–2015*.

Visser, H., Petersen, A. C., & Ligtvoet, W. (2014). On the relation between weather-related disaster impacts, vulnerability and climate change. *Climatic Change*, *125*(3), 461-477. https://doi.org/10.1007/s10584-014-1179-z

Watts, N., Amann, M., Arnell, N., Ayeb-Karlsson, S., Belesova, K., Boykoff, M., Byass, P., Cai, W., Campbell-Lendrum, D., & Capstick, S. (2019). The 2019 report of The Lancet Countdown on health and climate change: ensuring that the health of a child born today is not defined by a changing climate. *The Lancet*, *394*(10211), 1836-1878. https://doi.org/10.1016/S0140-6736(19)32596-6

Weatherlogics. (2018). *Case Study - Southern Ontario Damaging Windstorm: 4 May 2018*.

Weinkle, J., Landsea, C., Collins, D., Musulin, R., Crompton, R. P., Klotzbach, P. J., & Pielke, R. (2018). Normalized hurricane damage in the continental United States 1900–2017. *Nature Sustainability*, *1*(12), 808-813. https://doi.org/10.1038/s41893-018-0165-2

WMO. (2011). *Guide to Climatological Practices*. World Meteorological Organization.

WMO. (2016). *Technical Regulations, Basic Documents No. 2, Volume I – General Meteorological Standards and Recommended Practices (WMO-No. 49). 2015 edition, updated in 2016*. World Meteorological Organization.

Yamin, L. E., Hurtado, A. I., Barbat, A. H., & Cardona, O. D. (2014). Seismic and wind vulnerability assessment for the GAR-13 global risk assessment. *International journal of disaster risk reduction*, *10*, 452-460. https://doi.org/10.1016/j.ijdrr.2014.05.007

Ye, Y., & Fang, W. (2018). Estimation of the compound hazard severity of tropical cyclones over coastal China during 1949–2011 with copula function. *Natural Hazards*, *93*(2), 887-903. https://doi.org/10.1007/s11069-018-3329-5

Zhang, Q., Wu, L., & Liu, Q. (2009). Tropical cyclone damages in China 1983–2006. *Bulletin of the American Meteorological Society*, *90*(4), 489-496. https://doi.org/10.1175/2008BAMS2631.1

Zhang, Y., Wei, K., Shen, Z., Bai, X., Lu, X., & Soares, C. G. (2020). Economic impact of typhoon-induced wind disasters on port operations: A case study of ports in China. *International journal of disaster risk reduction*, *50*, 101719. https://doi.org/10.1016/j.ijdrr.2020.101719